\documentclass[aps,reprint,superscriptaddress,footnote, twocolumn,notitlepage]{revtex4-1}

\usepackage{appendix}

\usepackage{tikz}  
\usetikzlibrary{arrows,shapes,positioning,shadows,backgrounds,fit}

\usepackage{relsize}
\usepackage{graphicx}
\usepackage{amsmath, bbm}
\usepackage{amssymb}
\usepackage{amsthm,mathtools}
\usepackage{mathrsfs}
\usepackage{xcolor}
\usepackage{cancel}
\usepackage{titlesec}
\usepackage{enumitem}

\DeclareSymbolFontAlphabet{\amsmathbb}{AMSb}

\usepackage[normalem]{ulem}

\titlespacing{\section}{0ex}{2ex}{0.4ex}
\titleformat{name=\section}{\bf}{\thesection.}{.3em}{}

\usepackage{soul}
\usepackage{dsfont}

\def\be{\begin{eqnarray}}
\def\ee{\end{eqnarray}}
\newcommand{\tr}[1]{\text{Tr}\big(#1\big)}

\newcommand{\ket}[1]{|{#1}\rangle}
\newcommand{\bra}[1]{\langle{#1}|}

\renewcommand{\ln}[1]{\mathrm{ln} \left({#1} \right)}

\newcommand{\Tr}[1]{\text{tr}\left(#1\right)}

\theoremstyle{plain}

\definecolor{myblue}{rgb}{0.2,0.2,0.8}
\definecolor{myblack}{rgb}{0,0,0}
\definecolor{myurl}{rgb}{0.1,0.1,0.4}

\usepackage[colorlinks=true,citecolor=myblue,linkcolor=myblack,urlcolor=myurl]{hyperref}

\begin{document}

\title{Geometry of work fluctuations versus efficiency in microscopic thermal machines} 

\author{Harry J.~D. Miller}
\affiliation{Department of Physics and Astronomy, The University of Manchester, Manchester M13 9PL, UK.}

\author{Mohammad Mehboudi}
\affiliation{Max-Planck-Institut f\"ur Quantenoptik, D-85748 Garching, Germany}
\affiliation{ICFO-Institut de Ciencies Fotoniques, The Barcelona Institute of Science and Technology, 08860 Castelldefels (Barcelona), Spain}

\date{\today}

\begin{abstract}
     When engineering microscopic machines, increasing efficiency can often come at a price of reduced reliability due to the impact of stochastic fluctuations. Here we develop a general method for performing multi-objective optimisation of efficiency and work fluctuations in thermal machines operating close to equilibrium in either the classical or quantum regime. Our method utilises techniques from thermodynamic geometry, whereby we match optimal solutions to protocols parameterised by their thermodynamic length. We characterise the optimal protocols for continuous-variable Gaussian machines, which form a crucial class in the study of thermodynamics for microscopic systems.
\end{abstract}

\maketitle

Designing optimal protocols for heat-to-work conversion below the nanoscale remains an ongoing challenge in the fields of stochastic and quantum thermodynamics \cite{Seifert2012,Benenti,Kosloff2013b}. For microscopic machines, efficiency and power alone are not the only figures of merit due to the additional influence of stochastic fluctuations. While a machine may extract work efficiently on average, it may be subject to large work fluctuations which hampers any reliability. Considerable effort is now being devoted to study this interplay between efficiency, power and reliability in small scale systems \cite{Funo2018a,Barato,Barato2016,Pietzonka2018,Holubec2018,Solon2018a,Horowitza,Guarnieri2019b,Abiuso2020,Denzler2020}. 

Dissipation can be minimised for general far-from-equilibrium processes using methods from optimal control theory \cite{Muratore-ginanneschi2011,Schmiedl,Zulkowski2014,Cavina,Deffner2018a,Vu2020}.  However finding solutions with such approaches is often limited to simple systems with few degrees of freedom, which makes it difficult to identify general design principles for efficient thermal machines beyond specific models.  On the other hand, a general method for optimising the efficiencies of machines operating \textit{close to equilibrium} was recently proposed by Brandner and Saito \cite{Brandner2020}. This method, applicable to both classical and quantum periodic heat engines, relies on expressing the engine's entropy production in terms of a metric over the Riemann manifold of equilibrium states of the working system. One can maximise the efficiency for any given protocol by reparameterising it in terms of the so-called \textit{thermodynamic length} \cite{Weinhold,weinholdMetricGeometryEquilibrium1975b,Ruppeiner1979,Salamon1983a,Schlogl1985,Ruppeiner,Crooks2007,Sivak2012a,Zulkowski2012,Scandi,Deffner}, which provides a measure of distance between configurations in the equilibrium manifold. The benefit of this approach is its simplicity; optimisation is achieved by a straightforward computation of the thermodynamic metric tensor which depends only on the equilibrium and relaxation properties of the machine \cite{Brandner2020}. 

While thermodynamic length provides a systematic way of determining efficient protocols, it remains to be seen how increasing efficiency impacts the work fluctuations. For systems connected to a single fixed-temperature reservoir, initial investigations have explored the simultaneous optimisation of the average dissipated work required to drive a system from one state to another and the associated minimal work fluctuations \cite{Solon2018a,Miller2019}. As this is a multi-objective optimisation problem, one must consider the boundary of allowed protocols where dissipation cannot be reduced any further without suffering an increase in fluctuations, or vice versa. This boundary, known as a Pareto front \cite{Miettinen1999}, only converges to a single point in regimes where the fluctuation-dissipation relation holds true \cite{Jarzynski1997d,Speck,Mandal2016a}, in which case there exists a unique protocol with minimal average dissipated work and work variance. At present, Pareto optimisation has not been analysed in the context of periodic heat engines operating between different temperatures. In this situation at least two figures of merit are thermodynamic efficiency versus the resulting work fluctuations, whose optimal protocols are not expected to coincide for both classical and quantum systems. This is due to the fact that efficiency and work fluctuations are typically not monotonically related to each other, which can prohibit the existence of a unique optimal protocol.

In this paper, we outline a general method for finding Pareto-optimal protocols interpolating between maximum efficiency and minimal work fluctuations for engines operating close to equilibrium. Remarkably, we show that such protocols can be found by constructing a new form of thermodynamic metric tensor and corresponding length. By parameterising any given protocol in terms of this generalised thermodynamic length, one may identify regimes of optimal efficiency and reliability in a straightforward manner. We illustrate our approach for quantum heat engines operating along discrete step-equilibration cycles \cite{Nulton2013,Anders2013a,Large2019,Scandi2019}, though our results also apply to classical heat engines and open quantum systems undergoing Lindblad dynamics. As a core application, we derive analytic expressions for thermodynamic length in general continuous variable Gaussian quantum systems \cite{Weedbrook}. Such systems form a major platform for studying thermodynamic processes in the microscopic regime, and our approach can be used to optimise any Gaussian thermal machine by using the natural tools in the Gaussian formalism that operate on the steady-state covariance matrix. As an example we determine Pareto fronts of optimal efficiency and reliability for a system of coupled harmonic oscillators driven by periodic changes in temperature, frequency and coupling strength.



Let's begin by considering a quantum system weakly coupled to a thermal environment at inverse temperature $\beta=1/k_{B}T$. The system is subject to external control via a set of $d$ mechanical parameters $\vec{\lambda}:=(\lambda^1,\lambda^2, ...\lambda^d)$ and we denote its Hamiltonian by $H(\vec{\lambda})$. In addition, we allow for external modulation of the environmental temperature. Collectively these set of variables define a cycle via a closed curve $\gamma:t\mapsto \vec{\Lambda}_t$ in the parameter space containing the vectors
\begin{align}
    \vec{\Lambda}:=\big\{\beta,\vec{\lambda} \big\}\in\mathbb{R}^{d+1},
\end{align}
and we label  $\Lambda^0=\beta$ and $\Lambda^{j}=\lambda^j$ for $j\geq 1$. The parameter space defines a manifold of equilibrium states, defined by $\pi(\vec{\Lambda}):=\exp{(-\beta H(\vec{\lambda}))}/\tr{\exp{(-\beta H(\vec{\lambda}))}}$. Furthermore, we will introduce the following conjugate forces, 
\begin{align}\label{eq:X_i_def}
    X_j(\vec{\Lambda}):=\begin{cases}
    \beta^{-1}H(\vec{\lambda}), \ \ \ \ \ \ \ \ \   \text{if} \ \ \ j=0, \\
    \frac{\partial}{\partial \Lambda^j}H(\vec{\lambda}), \ \ \ \ \ \ \ \ \ \text{if} \ \ \ j\geq 1. \\
    \end{cases}
\end{align}
During the cycle the inverse temperature undergoes a variation between a maximum and minimum, $\beta_h\leq \beta(t)\leq\beta_c$, which we will express in the form $\beta(t):=\beta_c+(\beta_h-\beta_c)\delta\beta(t)$ with $\delta\beta(t)\in[0,1]$ a dimensionless periodic function. For convenience we take $t\in[0,1]$ to be a dimensionless parameter, and denote a discretised set of $N$ points along the curve evaluated at times $t_n=(n-1)/(N-1)$ for $n\in[1,N]$.  We have in mind thermodynamic cycles where each step can be approximated by a fast quench in the mechanical parameters $\vec{\lambda}_{t_n}\to\vec{\lambda}_{t_{n+1}}$, followed by relaxation to a new temperature $\beta_{t_n}\to \beta_{t_{n+1}}$. This means that at the beginning of each n'th step the system is in a thermal state $\pi(\vec{\Lambda}_{t_n})$, which is left unchanged during the quench step while work is performed. The state after the quench then relaxes to a new equilibrium state $\pi(\vec{\Lambda}_{t_{n+1}})$ with no work done during this step.   

A central quantity of interest is the average irreversible entropy production along the cycle $\gamma$, which can be expressed as 
\begin{align}\label{eq:entprod}
    S_{irr}&:=\beta_c W+(\beta_c-\beta_h)Q_{\text{in}} =\sum^{N-1}_{n=1}S\big(\pi(\vec{\Lambda}_{t_{n}})||\pi(\vec{\Lambda}_{t_{n+1}})\big),
\end{align}
where $S(\rho||\rho')=\tr{\rho \ \ln{\rho}}-\tr{\rho \ \ln{\rho'}}\geq 0$ the quantum relative entropy, and we identify the work done 
\begin{align}
    W:=\sum^{N-1}_{n=1} \tr{\big(H(\vec{\lambda}_{t_{n+1}})-H(\vec{\lambda}_{t_{n}})\big)\pi(\vec{\Lambda}_{t_{n}})}
\end{align}
and supplied heat
\begin{align}
    Q_{\text{in}}:=\sum^{N-1}_{n=1} \delta \beta(t_{n+1})\tr{H(\vec{\lambda}_{t_n})\big(\pi(\vec{\Lambda}_{t_{n+1}})-\pi(\vec{\Lambda}_{t_{n}})\big)}.
\end{align}
The second equality in~\eqref{eq:entprod} follows from using the periodicity $\vec{\Lambda}_{t_1}=\vec{\Lambda}_{t_N}$ (see Appendices A and B). This formula motivates a definition of efficiency for processes with a positive work output, $W\leq 0$, given by the ratio \cite{Brandner2016a,Brandner2018}
\begin{align}
    \eta:=-\frac{W}{Q_{\text{in}}}\leq \eta_C,
\end{align}
where $\eta_C=1-\beta_h/\beta_c$ denotes the Carnot efficiency. Here we see consistency with Carnot's theorem, which follows as a consequence of the second law $S_{irr}\geq 0$. 

\begin{figure*}
    \begin{tabular}{c}
    \includegraphics[width=\columnwidth]{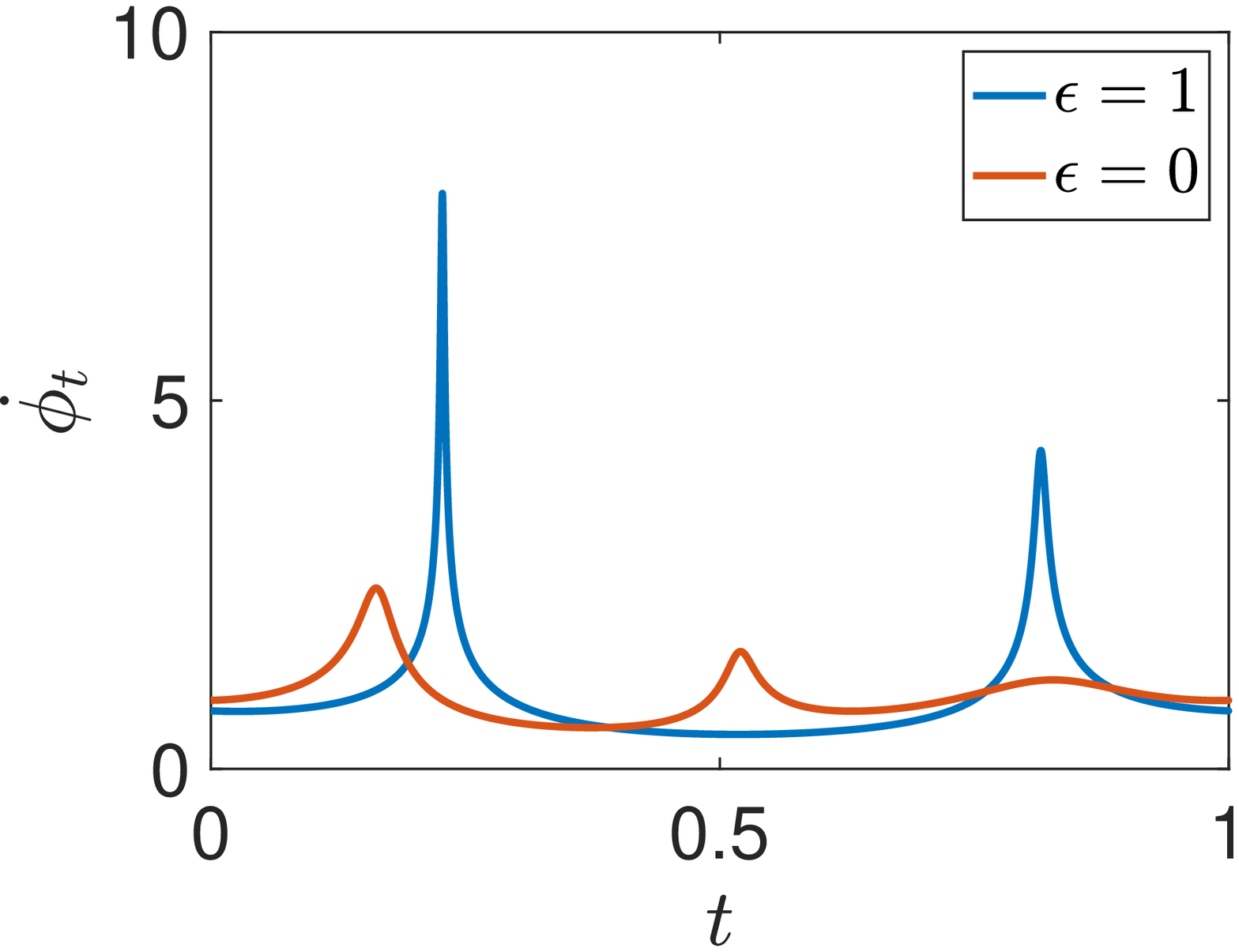}
    \includegraphics[width=\columnwidth]{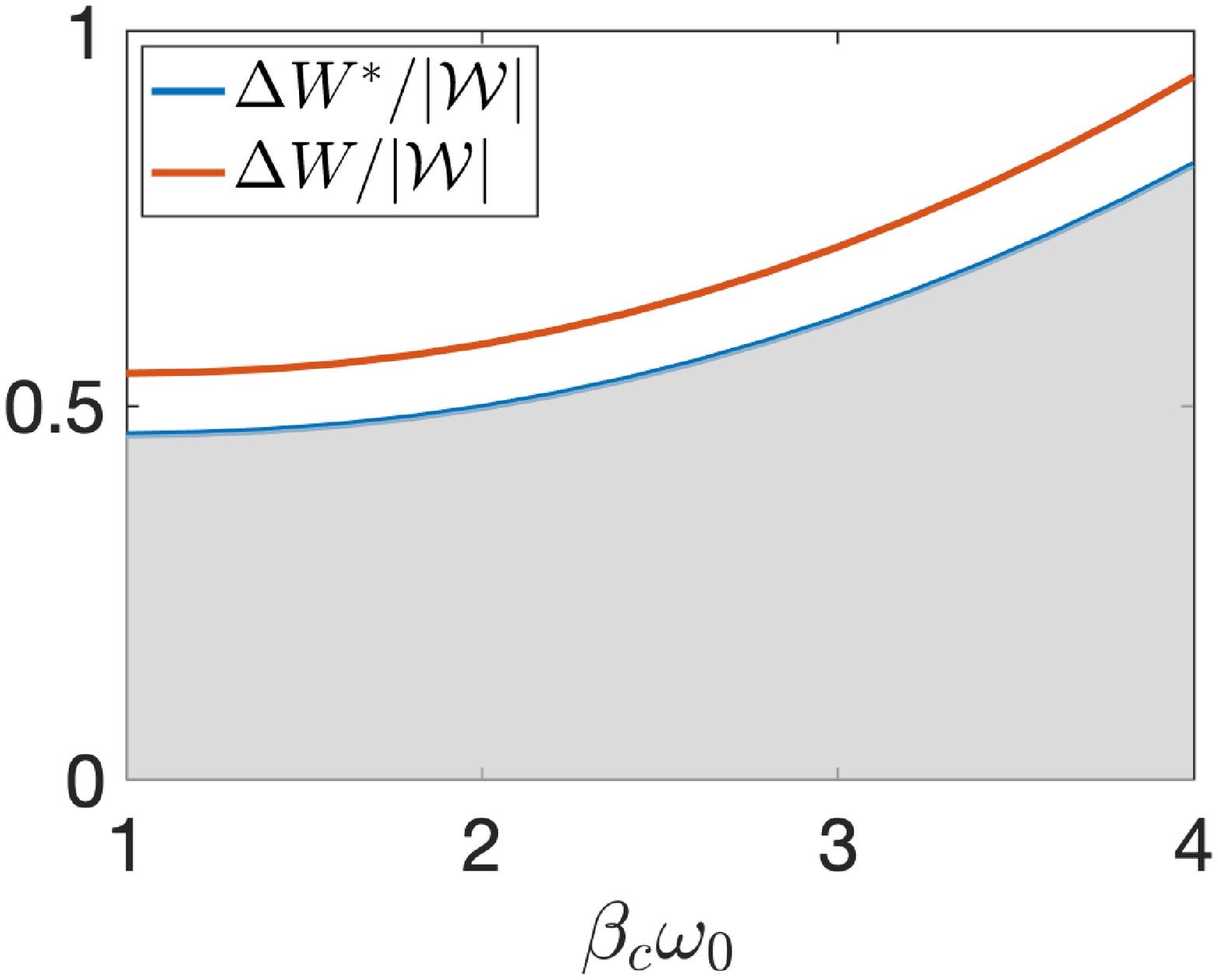}
    \end{tabular}
    \caption{\textbf{Left}---Derivative of the optimal speed $\phi^\epsilon_t$~\eqref{eq:parameter}, corresponding to work fluctuations optimisation at $\epsilon=1$ (the blue curve) and efficiency optimisation with $\epsilon=0$ (the red curve). The parameters are set to $\omega_0=1$, $T_c = 0.25\omega_0$, $T_h = T_c + \Delta T$ with $\Delta T = \omega_0$, $\kappa = 0.4\omega_0$. \textbf{Right}---Work fluctuations vs. oscillator frequency $\omega_0$, for the linear protocol (red) and the optimal protocol (blue). We denote $\Delta W=\sqrt{\text{Var}(W)}$ as the standard deviation and $\Delta W^*=k_B T_c\mathcal{L}_1/\sqrt{N}$ the optimal fluctuations given by~\eqref{eq:workbound}. Both quantities are plotted relative to the adiabatic work extracted $|\mathcal{W}|$, which is independent of parameterisation. The gray area is not accessible through any protocol. Here we set the parameters to $N=50$, $T_c=0.25$, $T_h = T_c + \Delta T$ with $\Delta T = 4T_c$, and $\kappa = 0.4T_c$. }
    \label{fig:my_label}
\end{figure*}

In addition to efficiency, we will also be concerned with the amount of work fluctuations generated along the cycle. For quantum systems, stochastic work is determined from projective measurements of the system energy at the beginning and end of each unitary step \cite{Talkner2007c}. By summing up the changes in energy across each step and computing the variance from the resulting work probability distribution, one can show \cite{Miller2019,Scandi2019}
\begin{align}\label{eq:var}
    \nonumber\text{Var}(W):=\sum^{N-1}_{n=1}&\tr{\big(H(\vec{\lambda}_{t_{n+1}})-H(\vec{\lambda}_{t_{n}})\big)^2\pi(\vec{\Lambda}_{t_n})} \\
    &-\tr{\big(H(\vec{\lambda}_{t_{n+1}}) 
    -H(\vec{\lambda}_{t_{n}})\big)\pi(\vec{\Lambda}_{t_n})}^2,
\end{align}
At this stage, we restrict our attention to cycles composed of a large number of steps $N^2\gg 1$, which defines a regime that is close to quasi-static \cite{Nulton2013}. In this case we may replace the summation with an integral over the continuous path $\gamma$ and obtain 
\begin{align}\label{eq:variance}
    \text{Var}(W)\simeq\frac{1}{N}\int_\gamma dt \  m_{jk}(\vec{\Lambda}_t)\frac{d\lambda^j}{dt}\frac{d\lambda^k}{dt}, 
\end{align}
where summation is carried out over repeated indices, restricted to the mechanical variables only. Here $m_{jk}$ is a positive semi-definite and symmetric tensor
\begin{align}\label{eq:metricfluc}
    m_{jk}(\vec{\Lambda}):=
    \frac{1}{2}\tr{\pi(\vec{\Lambda}) \big\{\delta X_j(\vec{\Lambda}), \delta X_k(\vec{\Lambda}) \big\}}, \ \ \ \text{if} \ \ \ \ \ j,k\geq 1 
    \end{align}
and we set $m_{j0}=m_{0j}=0 \ \ \forall j$ for later convenience. We also denote the shifted operators $\delta X_j (\vec{\Lambda})=X_j(\vec{\Lambda})-\tr{X_j(\vec{\Lambda}) \pi(\vec{\Lambda})}$ and $\{,\}$ represents the anti-commutator. Turning to the efficiency, we consider the fraction of the efficiency below the Carnot value, defined as $\delta \eta:=1-\frac{\eta}{\eta_C}$. Assuming a large number of steps and applying a Taylor expansion in $1/N$ to~\eqref{eq:entprod} gives
\begin{align}\label{eq:eta}
    \delta \eta \simeq -\frac{1}{2N \beta_c \mathcal{W}}\int_\gamma dt \  \ g_{jk}(\vec{\Lambda}_t)\frac{d\Lambda^j}{dt}\frac{d\Lambda^k}{dt}, 
\end{align}
where we now have a different metric tensor:
\begin{align}\label{eq:metric_efficiency}
    g_{jk}(\vec{\Lambda}):=\beta \int^\beta_0 dx \ \tr{\pi(\vec{\Lambda}) \ \delta X_j(\vec{\lambda}) \mathscr{U}_{ix,\vec{\lambda}}\big[\delta X_k(\vec{\Lambda})\big]},
\end{align}
and we have introduced the unitary channel $\mathscr{U}_{\nu,\vec{\lambda}} \ [.]=e^{i\nu H(\vec{\lambda})}[.]e^{-i\nu H(\vec{\lambda})}$. We have further defined the \textit{adiabatic} work done \cite{Brandner2020}:
\begin{align}\label{eq_adiabatic_work}
    \mathcal{W}:=\oint_\gamma \tr{X_j(\vec{\Lambda}) \pi(\vec{\Lambda})}d\lambda^j,
\end{align}
which is a geometric quantity independent of the parameterisation and assumed negative $\mathcal{W}\leq 0$ to ensure a useful work extraction cycle. We provide a proof of~\eqref{eq:metricfluc} and~\eqref{eq:metric_efficiency} in Appendix B. The tensor~\eqref{eq:metric_efficiency} is proportional to Kubo-Mori Fisher information metric \cite{Petz1994a,Hayashi2002}, which is a quantum analogue of the classical Fisher-Rao metric. At this stage we observe that the elements of this tensor typically differ from $m_{jk}$ due to possible non-commutativity between the conjugate forces, ie. if $[X_j,X_k]\neq 0$. In Appendix A we highlight the different information-geometric interpretations of these two metrics. 

To establish our optimisation problem, let us introduce the dimensionless multi-objective function
\begin{align}\label{eq:pareto}
    \mathcal{I}_\epsilon:=\epsilon  \text{Var}(\tilde{W})+(1-\epsilon)\delta\eta, \ \ \ \ \epsilon\in[0,1],
\end{align}
where for convenience we have defined the work fluctuations in units of the cold temperature, $\text{Var}(\tilde{W})=\beta_c^2\text{Var} (W)$. The question we now address is: how long should the system spend at each point along the protocol $\gamma$ in order to maximise efficiency while minimising fluctuations? This amounts to finding the best choice of parameterisation $\gamma: t\to \vec{\Lambda}_{t}'=\vec{\Lambda}_{\phi^\epsilon_t}$ with function $\phi^\epsilon_t\in[0,1]$ to be determined so as to minimise the scalarized objective~\eqref{eq:pareto}. Optimal parameterisations $\mathcal{I}^*_\epsilon\leq\mathcal{I}_\epsilon$ lie along sections of the Pareto fronts \cite{Miettinen1999}; these points form the boundary of protocols where it is not possible to increase efficiency (ie. reduce $\delta \eta$) without increasing work fluctuations, or conversely, reduce fluctuations without reducing efficiency. By combining~\eqref{eq:variance} with~\eqref{eq:eta} and applying the Cauchy-Schwarz inequality we arrive a geometric expression for the minimised objective function:
\begin{align}
    \mathcal{I}_\epsilon^*=\frac{\mathcal{L}_\epsilon^2}{N}, \ \ \ \ \ \ \ \text{where} \ \ \ \ \  \mathcal{L}_\epsilon=\oint_\gamma \sqrt{ M_{jk}(\vec{\Lambda})  \ d\Lambda^j d\Lambda^k},
\end{align}
and we define
\begin{align}\label{eq:metriceps}
    M_{jk}^\epsilon(\vec{\Lambda}):=\epsilon \beta_c^2 \  m_{jk}(\vec{\Lambda})+\frac{(1-\epsilon)}{2\beta_c |\mathcal{W}|}g_{jk}(\vec{\Lambda}).
\end{align}
This follows from the fact that $M_{jk}^\epsilon(\vec{\Lambda})$ gives another metric tensor, since it is formed by a positive-weighted linear combination of two metric tensors for $\epsilon\in[0,1]$. The function $\mathcal{L}_\epsilon$ may be interpreted as a generalised form of \textit{thermodynamic length} \cite{Crooks2007}, whose dependence on $\epsilon$ encodes information about the Pareto optimal solution. In the classical regime where we may describe the thermal system by a probability distribution $p(\vec{\Lambda})$, we may determine the equivalent tensor from the Fisher-Rao metric $F_{jk}(\vec{\Lambda})=\big \langle \partial_{\Lambda_j}\text{ln} \ p(\vec{\Lambda})  \ \partial_{\Lambda_k}\text{ln} \ p(\vec{\Lambda}) \big\rangle $ (see Appendix A). In this case~\eqref{eq:metriceps} takes the form
\begin{align}\label{eq:metric_class}
    M_{jk}^\epsilon(\vec{\Lambda})=\bigg(\epsilon \bigg( \frac{T}{T_c}\bigg)^2 \mu_{jk}+\frac{(1-\epsilon)}{2\beta_c |\mathcal{W}|}\bigg) F_{jk}(\vec{\Lambda}),
\end{align}
with $\mu_{j0}=\mu_{0k}=0 \ \forall j,k$ and $\mu_{jk}=1 \ \forall j,k>0$.

Crucially, these Pareto optimal solutions are determined by parameterising the protocol $\gamma$ in terms of the modified thermodynamic length via the speed function $t\to\phi_t^\epsilon$ \footnote{In this case one has $t\propto s(\phi_t^\epsilon)$, where $s(\phi_t^\epsilon)$ is the arc length for the interval $[0,\phi_t^\epsilon]$. This means the curve $\gamma:t\mapsto \vec{\Lambda}_t$ is traversed at constant velocity, leading to the equality condition for the Cauchy-Schwartz inequality. }, which is obtained from the implicit equation:
\begin{align}\label{eq:parameter}
    t=\frac{1}{\mathcal{L}_\epsilon}\int^{\phi^\epsilon_t}_0 ds \  \sqrt{M^\epsilon_{jk}(\vec{\Lambda}_s)\frac{d\Lambda^j}{ds}\frac{d\Lambda^k}{ds}}, 
\end{align}
This means that for any given protocol, optimisation is achieved by changing the speed at which the curve is traversed by choosing a new parameterisation $\gamma:t\mapsto \vec{\Lambda}_{t}'=\vec{\Lambda}_{\phi^\epsilon_t}$. As long as the solution satisfies $(|\dot{\phi}_t|/N)^2\ll 1$ at all times, the system remains sufficiently close to equilibrium and the optimal protocols may be realised. There are two limiting cases of the geometric bound. For $\epsilon=1$ we obtain a geometric lower bound on the achievable work fluctuations: 
\begin{align}\label{eq:workbound}
    \text{Var}(W)\geq \frac{(k_B T_c)^2}{N}\mathcal{L}_1^2.
\end{align}
For $\epsilon=0$, we obtain a maximum upper bound on efficiency $\eta\leq \eta_C(1-\mathcal{L}_0^2/N)$. An analogous efficiency bound was previously obtained in~\cite{Brandner2020} for continuous Lindblad dynamics.

\begin{figure*}[t!]
    \begin{tabular}{c}
    \includegraphics[width=\columnwidth]{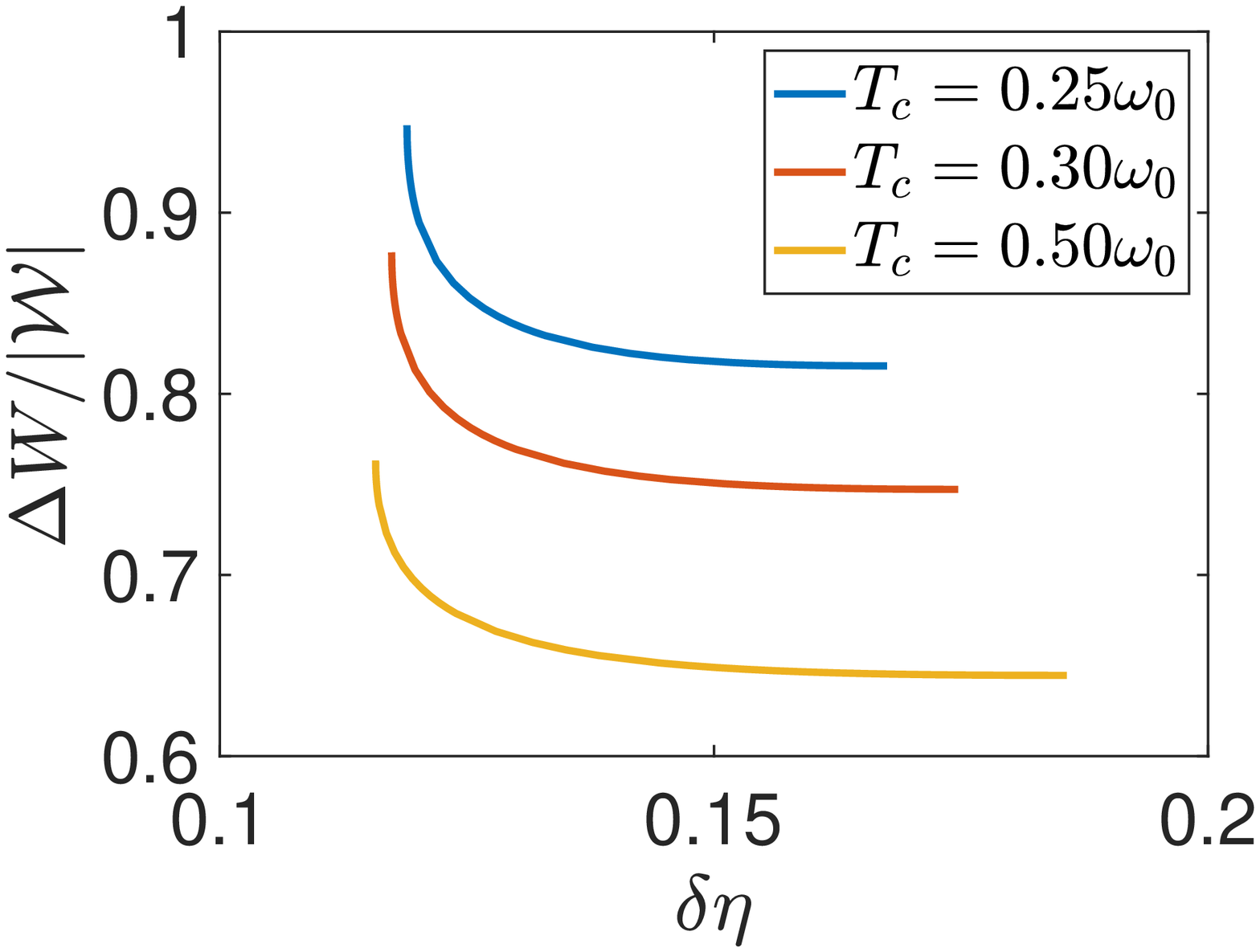}
    \includegraphics[width=\columnwidth]{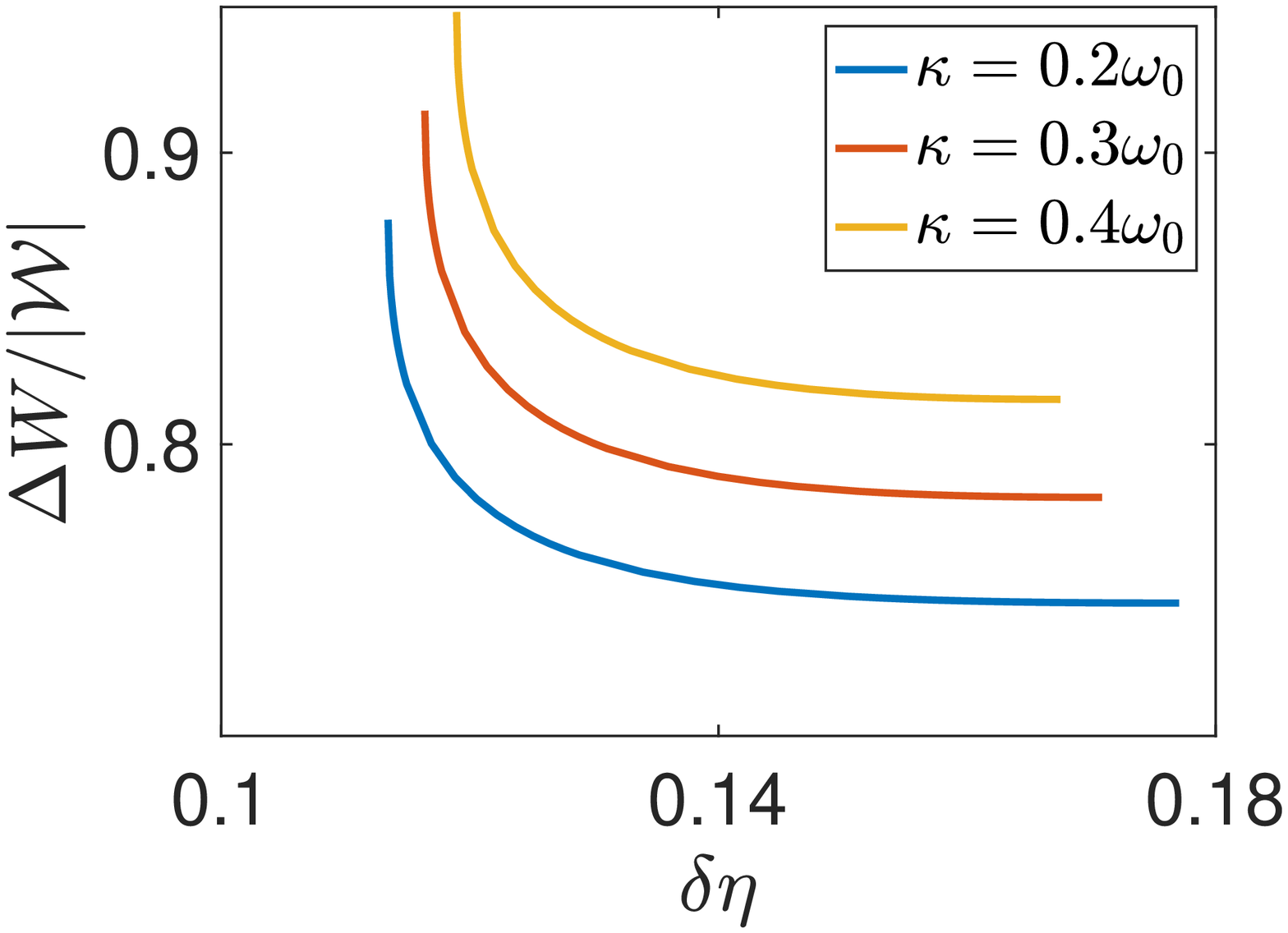}
    \end{tabular}
    \caption{\textbf{Left}---The \textit{relative work-fluctuations vs. efficiency} Pareto fronts for different values of cold temperature $T_c$. Overall neither the relative work fluctuations nor the efficiency have a monotonic behaviour with temperature. The parameters are set to  $\omega_0=2$, $T_h = T_c + \Delta T$ with $\Delta T = \omega_0$, $\kappa = 0.4\omega_0$. \textbf{Right}---Same as left, for different couplings between the two oscillators. Here we set $\omega_0 = 1$, $T_c = 0.25\omega_0$, $T_h = T_c + \Delta T$ with $\Delta T = \omega_0$ and $N=50$.}
    \label{fig:my_label2}
\end{figure*}

The above construction gives a general recipe for finding Pareto optimal protocols for arbitrary quantum or classical systems, valid in regimes where the number of steps between equilibrium states is large. A particular class of  systems that are frequently used to describe many relevant physical systems in thermodynamics, such as ion trap heat engines \cite{Rossnagel2015},  are composed of Gaussian quantum states \cite{Weedbrook,Mehboudi}. The Hamiltonian of a \textit{D}-mode Gaussian system is quadratic in quadrature operators, namely
\begin{align}
    H(\vec{\lambda}) = \frac{1}{2}R^T\mathbb{G}_{\vec{\lambda}}R,
\end{align}
with $R=(x_1,~p_1,\dots, x_D,~p_D)^T$ being the quadrature vector. Here, the $D\times D$ dimensional \textit{symmetric} matrix $\mathbb{G}_{\vec{\lambda}}$ contains all the quadratic couplings. For this general class we provide an analytic formula for the Pareto optimal solutions, which are given by~\eqref{eq:parameter} via computing the metric tensor~\eqref{eq:metriceps}. By defining $\mathbb{X}_j \coloneqq \partial_{\Lambda_j}\mathbb{G}_{\vec{\lambda}}$ if $j\geq 1$ and $\mathbb{X}_0 \coloneqq \beta^{-1} \mathbb{G}_{\vec{\lambda}}$ we can express the metric~\eqref{eq:metriceps} as follows (see Appendix D):
\begin{align}\label{eq:metricGauss}
    \nonumber M_{jk}^\epsilon(\vec{\Lambda})  = a_{jk}\frac{\epsilon \beta_c^2}{4}\big\{ \Tr{ \mathbb{X}_j \underline {\mathbb{X}}_k} + \Tr{ \mathbb{X}_k \underline {\mathbb{X}}_j} \big\}  \\
     +\frac{(1-\epsilon)\beta}{4\beta_c |\mathcal{W}|}\Tr{{\bar {\mathbb{X}}}_j \ \underline{\mathbb{X}}_k},
\end{align}
where $a_{j0}=a_{0j}=0 \ \ \forall j$ and $a_{jk}=1 \ \ \forall j,k>0$, and we define
\begin{align}
    \bar{\mathbb{X}}_j & = \int_0^{\beta}dx~ [e^{ix\Omega \mathbb{G}_{\vec{\lambda}}}]^T \mathbb{X}_j [e^{ix\Omega \mathbb{G}_{\vec{\lambda}}}],\label{eq:Gauss_X_1}\\
    \underline {\mathbb{X}}_j & = (\sigma(\vec{\Lambda}) - \frac{1}{2}\Omega)  \mathbb{X}_j(\sigma(\vec{\Lambda}) + \frac{1}{2}\Omega)\label{eq:Gauss_X_2},
\end{align}
with $\Omega$ being the symplectic form with $\Omega_{nm} = i[R_n,R_m]$, and $\sigma(\vec{\Lambda})$ representing the steady state covariance matrix with elements $[\sigma(\vec{\Lambda})]_{nm} = \tr{\pi(\vec{\Lambda})\{R_n,R_m\}}/2-\tr{\pi(\vec{\Lambda})R_n}\tr{\pi(\vec{\Lambda})R_m}$. Furthermore, the adiabatic work is found using $\mathcal{W}=1/2\oint_\gamma d\lambda^j{\rm tr}\left({\mathbb{X}_j \sigma(\vec{\Lambda}})\right)$. Notice that the `${\text{tr}}$' operation acts as a trace on the matrix space associated to the Gaussian covariance matrices, which should be distinguished from the trace `${\text{Tr}}$' which acts on the Hilbert space for density operators.

We have now derived the general form for the thermodynamic metric tensor for Gaussian heat engines. So long as one can compute this metric tensor, the speed function $\phi_t^\epsilon$ can be approximately determined from~\eqref{eq:parameter} via point-wise inversion followed by numerical interpolation. We illustrate our method for the example of a pair of coupled harmonic oscillators with $R = [x_1~ p_1~ x_2~ p_2]^T$ and Hamiltonian coefficient matrix
\begin{align}\label{eq:G_Matrix}
    \mathbb{G}_{\vec{\lambda}} = \left[
    \begin{array}{cccc}
        \omega^2+\kappa & 0 & -\kappa & 0 \\
        0 & 1 & 0 & 0\\
        -\kappa & 0 & \omega^2+\kappa & 0\\
        0 & 0 & 0 & 1\\
    \end{array}
    \right],
\end{align}
were we chose equal frequencies $\omega_1=\omega_2=\omega$ and denote $\kappa$ the coupling strength between the oscillators. As for the driving protocol, we consider control over the bath temperature alongside the joint frequency and coupling, ie. $\gamma: t\mapsto \vec{\Lambda}_t=\{\beta(t), \omega(t),\kappa(t)\}$. The matrices ${\mathbb X}_{\omega} = \partial_{\omega} \mathbb{G}_{\vec{\lambda}}$, ${\mathbb X}_{\kappa} = \partial_{\kappa} \mathbb{G}_{\vec{\lambda}}$, and ${\mathbb X}_{0} = \beta^{-1} \mathbb{G}_{\vec{\lambda}}$ are easily found from Eq.~\eqref{eq:G_Matrix}. By substituting in Eqs.~\eqref{eq:Gauss_X_1} and \eqref{eq:Gauss_X_2} one finds the corresponding $\bar{ \mathbb{X}}_j$ and $\underline{ \mathbb{X}}_j$. Finally by plugging these into Eq.~\eqref{eq:metricGauss} we find the metric \footnote{The metric has a long and cumbersome analytical expression that we do not present.}. We choose a harmonic protocol path $\beta(t)=\beta_c+(\beta_h-\beta_c)\text{sin}^2(\pi t)$, $\omega(t)=\omega_0\left(1+\text{sin}^2(\pi t+\frac{\pi}{4})\right)$, and $\kappa(t)=\kappa_0\left(1+\text{sin}^2(\pi t+\frac{\pi}{4})\right)$, with the parameters $\kappa_0$, $\omega_0$, $\beta_c$ and $\beta_h$ being fixed during the cycle. In Figure~\ref{fig:my_label} we compare the work fluctuations $\Delta W=\sqrt{\text{Var}(W)}$ for a linear parameterisation as a function of the optimal amount $\Delta W^*=k_B T_c\mathcal{L}_1/\sqrt{N}$ given by~\eqref{eq:workbound}, both as a function of oscillator frequency $\omega_0$ and expressed in units of the adiabatic work. We also plot the rate of change in the speed function $\phi^{\epsilon}_t$ versus time for $\epsilon=1$, giving minimal fluctuations, compared with $\epsilon=0$ that gives maximum efficiency. One can clearly see that distinct protocols must be chosen in order to achieve either optimal efficiency or fluctuations. In Figure~\ref{fig:my_label2} we plot the Pareto fronts for the model for different choices of cold temperature $T_c$ and coupling constant $\kappa$. The points on each curve give the corresponding values of $\delta \eta$ and $\Delta W$ for the optimal protocol $\vec{\Lambda}_t'=\vec{\Lambda}_{\phi_t^\epsilon}$ determined from~\eqref{eq:parameter} for every $\epsilon\in[0,1]$. These curves form the boundary of achievable fluctuations and efficiency for the chosen protocol $\gamma$, with points below the curves inaccessible. In this case the Pareto fronts are strictly convex, and hence the entire front is determined by the minima of the scalarised objective function~\eqref{eq:pareto} \footnote{For a convex Pareto front, the full set of minima of $\mathcal{I}^*_\epsilon \ \ \forall \epsilon\in[0,1]$ characterise the entire front. However, points situated along any non-convex regions of the front cannot be determined from minimising $\mathcal{I}_\epsilon$ alone. In such situations, one can additionally apply an adaptive weighted sum method \cite{Kim2005} to determine any non-convex sections of the front. }. 

To summarise, we have constructed a general method for performing multi-objective optimisation of efficiency and work fluctuations in microscopic heat engines operating close to equilibrium. This method relies on determining a thermodynamic metric tensor that encodes information about both the efficiency and fluctuations simultaneously. While we have here focused on the quantum regime, our method can be readily applied to classical-stochastic systems using the tensor~\eqref{eq:metric_class}, which we demonstrate in Appendix C for the optimisation of a classical harmonic oscillator. This formalism opens an avenue to further applications and generalisations. For example in the Appendix we show that an analogous metric providing Pareto optimal solutions for work fluctuations and efficiency can be derived for engine cycles described using Lindblad dynamics, and we derive the corresponding expressions for Gaussian Lindbladians (see Appendix E). Future investigations could focus on extending our approach to strongly-coupled quantum heat engines \cite{Newman2017}, or regimes far away from equilibrium \cite{Deffner2018b,Vu2020}. The Gaussian metric tensors we have derived can also be used to study other aspects of thermodynamic geometry, such as computing scalar curvature \cite{Ruppeiner} and geodesics along the manifold of thermal Gaussian states \cite{Sivak2012a}. 

\acknowledgements
We thank Mart\'i Perarnau-Llobet for useful comments on the manuscript. This work was financially supported by Spanish MINECO (QIBEQI FIS2016-80773-P, ConTrAct FIS2017-83709-
R, and Severo Ochoa SEV-2015-0522), Fundacio Privada Cellex, and the Generalitat de Catalunya (CERCA Program and SGR1381). H. J. D. M. acknowledges support from the Royal Commission for the Exhibition of 1851.

\bibliographystyle{apsrev4-1}
\bibliography{mybib3.bib}

\begin{thebibliography}{62}%
\makeatletter
\providecommand \@ifxundefined [1]{%
 \@ifx{#1\undefined}
}%
\providecommand \@ifnum [1]{%
 \ifnum #1\expandafter \@firstoftwo
 \else \expandafter \@secondoftwo
 \fi
}%
\providecommand \@ifx [1]{%
 \ifx #1\expandafter \@firstoftwo
 \else \expandafter \@secondoftwo
 \fi
}%
\providecommand \natexlab [1]{#1}%
\providecommand \enquote  [1]{``#1''}%
\providecommand \bibnamefont  [1]{#1}%
\providecommand \bibfnamefont [1]{#1}%
\providecommand \citenamefont [1]{#1}%
\providecommand \href@noop [0]{\@secondoftwo}%
\providecommand \href [0]{\begingroup \@sanitize@url \@href}%
\providecommand \@href[1]{\@@startlink{#1}\@@href}%
\providecommand \@@href[1]{\endgroup#1\@@endlink}%
\providecommand \@sanitize@url [0]{\catcode `\\12\catcode `\$12\catcode
  `\&12\catcode `\#12\catcode `\^12\catcode `\_12\catcode `\%12\relax}%
\providecommand \@@startlink[1]{}%
\providecommand \@@endlink[0]{}%
\providecommand \url  [0]{\begingroup\@sanitize@url \@url }%
\providecommand \@url [1]{\endgroup\@href {#1}{\urlprefix }}%
\providecommand \urlprefix  [0]{URL }%
\providecommand \Eprint [0]{\href }%
\providecommand \doibase [0]{http://dx.doi.org/}%
\providecommand \selectlanguage [0]{\@gobble}%
\providecommand \bibinfo  [0]{\@secondoftwo}%
\providecommand \bibfield  [0]{\@secondoftwo}%
\providecommand \translation [1]{[#1]}%
\providecommand \BibitemOpen [0]{}%
\providecommand \bibitemStop [0]{}%
\providecommand \bibitemNoStop [0]{.\EOS\space}%
\providecommand \EOS [0]{\spacefactor3000\relax}%
\providecommand \BibitemShut  [1]{\csname bibitem#1\endcsname}%
\let\auto@bib@innerbib\@empty
\bibitem [{\citenamefont {Seifert}(2012)}]{Seifert2012}%
  \BibitemOpen
  \bibfield  {author} {\bibinfo {author} {\bibfnamefont {U.}~\bibnamefont
  {Seifert}},\ }\href@noop {} {\bibfield  {journal} {\bibinfo  {journal} {Rep.
  Prog. Phys}\ }\textbf {\bibinfo {volume} {75}},\ \bibinfo {pages} {126001}
  (\bibinfo {year} {2012})}\BibitemShut {NoStop}%
\bibitem [{\citenamefont {Benenti}\ \emph {et~al.}(2017)\citenamefont
  {Benenti}, \citenamefont {Casati}, \citenamefont {Saito},\ and\ \citenamefont
  {Whitney}}]{Benenti}%
  \BibitemOpen
  \bibfield  {author} {\bibinfo {author} {\bibfnamefont {G.}~\bibnamefont
  {Benenti}}, \bibinfo {author} {\bibfnamefont {G.}~\bibnamefont {Casati}},
  \bibinfo {author} {\bibfnamefont {K.}~\bibnamefont {Saito}}, \ and\ \bibinfo
  {author} {\bibfnamefont {R.~S.}\ \bibnamefont {Whitney}},\ }\href@noop {}
  {\bibfield  {journal} {\bibinfo  {journal} {Phys. Rep.}\ }\textbf {\bibinfo
  {volume} {1}},\ \bibinfo {pages} {694} (\bibinfo {year} {2017})}\BibitemShut
  {NoStop}%
\bibitem [{\citenamefont {Kosloff}\ and\ \citenamefont
  {Levy}(2013)}]{Kosloff2013b}%
  \BibitemOpen
  \bibfield  {author} {\bibinfo {author} {\bibfnamefont {R.}~\bibnamefont
  {Kosloff}}\ and\ \bibinfo {author} {\bibfnamefont {A.}~\bibnamefont {Levy}},\
  }\href {\doibase 10.1146/annurev-physchem-040513-103724} {\bibfield
  {journal} {\bibinfo  {journal} {Ann. Rev. Phys. Chem.}\ }\textbf {\bibinfo
  {volume} {65}},\ \bibinfo {pages} {365} (\bibinfo {year} {2013})}\BibitemShut
  {NoStop}%
\bibitem [{\citenamefont {Funo}\ and\ \citenamefont {Ueda}(2015)}]{Funo2018a}%
  \BibitemOpen
  \bibfield  {author} {\bibinfo {author} {\bibfnamefont {K.}~\bibnamefont
  {Funo}}\ and\ \bibinfo {author} {\bibfnamefont {M.}~\bibnamefont {Ueda}},\
  }\href@noop {} {\bibfield  {journal} {\bibinfo  {journal} {Phys. Rev. Lett.}\
  }\textbf {\bibinfo {volume} {115}},\ \bibinfo {pages} {260601} (\bibinfo
  {year} {2015})}\BibitemShut {NoStop}%
\bibitem [{\citenamefont {Barato}\ and\ \citenamefont
  {Seifert}(2015)}]{Barato}%
  \BibitemOpen
  \bibfield  {author} {\bibinfo {author} {\bibfnamefont {A.~C.}\ \bibnamefont
  {Barato}}\ and\ \bibinfo {author} {\bibfnamefont {U.}~\bibnamefont
  {Seifert}},\ }\href@noop {} {\bibfield  {journal} {\bibinfo  {journal} {Phys.
  Rev. Lett.}\ }\textbf {\bibinfo {volume} {114}},\ \bibinfo {pages} {158101}
  (\bibinfo {year} {2015})}\BibitemShut {NoStop}%
\bibitem [{\citenamefont {Barato}\ and\ \citenamefont
  {Seifert}(2016)}]{Barato2016}%
  \BibitemOpen
  \bibfield  {author} {\bibinfo {author} {\bibfnamefont {A.~C.}\ \bibnamefont
  {Barato}}\ and\ \bibinfo {author} {\bibfnamefont {U.}~\bibnamefont
  {Seifert}},\ }\href {\doibase 10.1103/PhysRevX.6.041053} {\bibfield
  {journal} {\bibinfo  {journal} {Phys. Rev. X}\ }\textbf {\bibinfo {volume}
  {6}},\ \bibinfo {pages} {041053} (\bibinfo {year} {2016})}\BibitemShut
  {NoStop}%
\bibitem [{\citenamefont {Pietzonka}\ and\ \citenamefont
  {Seifert}(2018)}]{Pietzonka2018}%
  \BibitemOpen
  \bibfield  {author} {\bibinfo {author} {\bibfnamefont {P.}~\bibnamefont
  {Pietzonka}}\ and\ \bibinfo {author} {\bibfnamefont {U.}~\bibnamefont
  {Seifert}},\ }\href@noop {} {\bibfield  {journal} {\bibinfo  {journal} {Phys.
  Rev. Lett.}\ }\textbf {\bibinfo {volume} {120}},\ \bibinfo {pages} {190602}
  (\bibinfo {year} {2018})}\BibitemShut {NoStop}%
\bibitem [{\citenamefont {Holubec}\ and\ \citenamefont
  {Ryabov}(2018)}]{Holubec2018}%
  \BibitemOpen
  \bibfield  {author} {\bibinfo {author} {\bibfnamefont {V.}~\bibnamefont
  {Holubec}}\ and\ \bibinfo {author} {\bibfnamefont {A.}~\bibnamefont
  {Ryabov}},\ }\href@noop {} {\bibfield  {journal} {\bibinfo  {journal} {Phys.
  Rev. Lett.}\ }\textbf {\bibinfo {volume} {120601}},\ \bibinfo {pages} {121}
  (\bibinfo {year} {2018})}\BibitemShut {NoStop}%
\bibitem [{\citenamefont {Solon}\ and\ \citenamefont
  {Horowitz}(2018)}]{Solon2018a}%
  \BibitemOpen
  \bibfield  {author} {\bibinfo {author} {\bibfnamefont {A.~P.}\ \bibnamefont
  {Solon}}\ and\ \bibinfo {author} {\bibfnamefont {J.~M.}\ \bibnamefont
  {Horowitz}},\ }\href@noop {} {\bibfield  {journal} {\bibinfo  {journal}
  {Phys. Rev. Lett.}\ }\textbf {\bibinfo {volume} {120}},\ \bibinfo {pages}
  {180605} (\bibinfo {year} {2018})}\BibitemShut {NoStop}%
\bibitem [{\citenamefont {Horowitz}\ and\ \citenamefont
  {Gingrich}(2020)}]{Horowitza}%
  \BibitemOpen
  \bibfield  {author} {\bibinfo {author} {\bibfnamefont {J.~M.}\ \bibnamefont
  {Horowitz}}\ and\ \bibinfo {author} {\bibfnamefont {T.~R.}\ \bibnamefont
  {Gingrich}},\ }\href {\doibase 10.1038/s41567-019-0702-6} {\bibfield
  {journal} {\bibinfo  {journal} {Nat. Phys.}\ }\textbf {\bibinfo {volume}
  {16}},\ \bibinfo {pages} {15} (\bibinfo {year} {2020})}\BibitemShut {NoStop}%
\bibitem [{\citenamefont {Guarnieri}\ \emph {et~al.}(2019)\citenamefont
  {Guarnieri}, \citenamefont {Landi}, \citenamefont {Clark},\ and\
  \citenamefont {Goold}}]{Guarnieri2019b}%
  \BibitemOpen
  \bibfield  {author} {\bibinfo {author} {\bibfnamefont {G.}~\bibnamefont
  {Guarnieri}}, \bibinfo {author} {\bibfnamefont {G.~T.}\ \bibnamefont
  {Landi}}, \bibinfo {author} {\bibfnamefont {S.~R.}\ \bibnamefont {Clark}}, \
  and\ \bibinfo {author} {\bibfnamefont {J.}~\bibnamefont {Goold}},\
  }\href@noop {} {\bibfield  {journal} {\bibinfo  {journal} {Phys. Rev.
  Research}\ }\textbf {\bibinfo {volume} {1}},\ \bibinfo {pages} {033021}
  (\bibinfo {year} {2019})}\BibitemShut {NoStop}%
\bibitem [{\citenamefont {Abiuso}\ and\ \citenamefont
  {Perarnau-Llobet}(2020)}]{Abiuso2020}%
  \BibitemOpen
  \bibfield  {author} {\bibinfo {author} {\bibfnamefont {P.}~\bibnamefont
  {Abiuso}}\ and\ \bibinfo {author} {\bibfnamefont {M.}~\bibnamefont
  {Perarnau-Llobet}},\ }\href@noop {} {\bibfield  {journal} {\bibinfo
  {journal} {Phys. Rev. Lett.}\ }\textbf {\bibinfo {volume} {124}},\ \bibinfo
  {pages} {110606} (\bibinfo {year} {2020})}\BibitemShut {NoStop}%
\bibitem [{\citenamefont {Denzler}\ and\ \citenamefont
  {Lutz}(2020)}]{Denzler2020}%
  \BibitemOpen
  \bibfield  {author} {\bibinfo {author} {\bibfnamefont {T.}~\bibnamefont
  {Denzler}}\ and\ \bibinfo {author} {\bibfnamefont {E.}~\bibnamefont {Lutz}},\
  }\href@noop {} {\  (\bibinfo {year} {2020})},\ \Eprint
  {http://arxiv.org/abs/arXiv:2007.01034} {arXiv:2007.01034} \BibitemShut
  {NoStop}%
\bibitem [{\citenamefont {Aurell}\ \emph {et~al.}(2011)\citenamefont {Aurell},
  \citenamefont {Mej{\'{i}}a-Monasterio},\ and\ \citenamefont
  {Muratore-Ginanneschi}}]{Muratore-ginanneschi2011}%
  \BibitemOpen
  \bibfield  {author} {\bibinfo {author} {\bibfnamefont {E.}~\bibnamefont
  {Aurell}}, \bibinfo {author} {\bibfnamefont {C.}~\bibnamefont
  {Mej{\'{i}}a-Monasterio}}, \ and\ \bibinfo {author} {\bibfnamefont
  {P.}~\bibnamefont {Muratore-Ginanneschi}},\ }\href {\doibase
  10.1103/PhysRevLett.106.250601} {\bibfield  {journal} {\bibinfo  {journal}
  {Phys. Rev. Lett.}\ }\textbf {\bibinfo {volume} {106}},\ \bibinfo {pages}
  {250601} (\bibinfo {year} {2011})}\BibitemShut {NoStop}%
\bibitem [{\citenamefont {Schmiedl}\ and\ \citenamefont
  {Seifert}()}]{Schmiedl}%
  \BibitemOpen
  \bibfield  {author} {\bibinfo {author} {\bibfnamefont {T.}~\bibnamefont
  {Schmiedl}}\ and\ \bibinfo {author} {\bibfnamefont {U.}~\bibnamefont
  {Seifert}},\ }\href@noop {} {\bibfield  {journal} {\bibinfo  {journal} {Phys.
  Rev. Lett.}\ }\textbf {\bibinfo {volume} {98}},\ \bibinfo {pages}
  {108301}}\BibitemShut {NoStop}%
\bibitem [{\citenamefont {Zulkowski}\ and\ \citenamefont
  {Deweese}(2014)}]{Zulkowski2014}%
  \BibitemOpen
  \bibfield  {author} {\bibinfo {author} {\bibfnamefont {P.~R.}\ \bibnamefont
  {Zulkowski}}\ and\ \bibinfo {author} {\bibfnamefont {M.~R.}\ \bibnamefont
  {Deweese}},\ }\href@noop {} {\bibfield  {journal} {\bibinfo  {journal} {Phys.
  Rev. E}\ }\textbf {\bibinfo {volume} {89}},\ \bibinfo {pages} {052140}
  (\bibinfo {year} {2014})}\BibitemShut {NoStop}%
\bibitem [{\citenamefont {Cavina}\ \emph {et~al.}(2018)\citenamefont {Cavina},
  \citenamefont {Mari}, \citenamefont {Carlini},\ and\ \citenamefont
  {Giovannetti}}]{Cavina}%
  \BibitemOpen
  \bibfield  {author} {\bibinfo {author} {\bibfnamefont {V.}~\bibnamefont
  {Cavina}}, \bibinfo {author} {\bibfnamefont {A.}~\bibnamefont {Mari}},
  \bibinfo {author} {\bibfnamefont {A.}~\bibnamefont {Carlini}}, \ and\
  \bibinfo {author} {\bibfnamefont {V.}~\bibnamefont {Giovannetti}},\
  }\href@noop {} {\bibfield  {journal} {\bibinfo  {journal} {Phys. Rev. A}\
  }\textbf {\bibinfo {volume} {98}},\ \bibinfo {pages} {012139} (\bibinfo
  {year} {2018})}\BibitemShut {NoStop}%
\bibitem [{\citenamefont {Bonanca}\ and\ \citenamefont
  {Deffner}(2018)}]{Deffner2018a}%
  \BibitemOpen
  \bibfield  {author} {\bibinfo {author} {\bibfnamefont {M.~V.~S.}\
  \bibnamefont {Bonanca}}\ and\ \bibinfo {author} {\bibfnamefont
  {S.}~\bibnamefont {Deffner}},\ }\href@noop {} {\bibfield  {journal} {\bibinfo
   {journal} {Phys. Rev. E}\ }\textbf {\bibinfo {volume} {98}},\ \bibinfo
  {pages} {042103} (\bibinfo {year} {2018})}\BibitemShut {NoStop}%
\bibitem [{\citenamefont {Vu}\ and\ \citenamefont {Hasegawa}(2020)}]{Vu2020}%
  \BibitemOpen
  \bibfield  {author} {\bibinfo {author} {\bibfnamefont {T.~V.}\ \bibnamefont
  {Vu}}\ and\ \bibinfo {author} {\bibfnamefont {Y.}~\bibnamefont {Hasegawa}},\
  }\href@noop {} {\  (\bibinfo {year} {2020})},\ \Eprint
  {http://arxiv.org/abs/arXiv:2005.02871} {arXiv:2005.02871} \BibitemShut
  {NoStop}%
\bibitem [{\citenamefont {Brandner}\ and\ \citenamefont
  {Saito}(2020)}]{Brandner2020}%
  \BibitemOpen
  \bibfield  {author} {\bibinfo {author} {\bibfnamefont {K.}~\bibnamefont
  {Brandner}}\ and\ \bibinfo {author} {\bibfnamefont {K.}~\bibnamefont
  {Saito}},\ }\href@noop {} {\bibfield  {journal} {\bibinfo  {journal} {Phys.
  Rev. Lett.}\ }\textbf {\bibinfo {volume} {124}},\ \bibinfo {pages} {040602}
  (\bibinfo {year} {2020})}\BibitemShut {NoStop}%
\bibitem [{\citenamefont {Weinhold}(1975{\natexlab{a}})}]{Weinhold}%
  \BibitemOpen
  \bibfield  {author} {\bibinfo {author} {\bibfnamefont {F.}~\bibnamefont
  {Weinhold}},\ }\href {\doibase 10.1063/1.431689} {\bibfield  {journal}
  {\bibinfo  {journal} {J. Chem. Phys.}\ }\textbf {\bibinfo {volume} {63}},\
  \bibinfo {pages} {2479} (\bibinfo {year} {1975}{\natexlab{a}})}\BibitemShut
  {NoStop}%
\bibitem [{\citenamefont
  {Weinhold}(1975{\natexlab{b}})}]{weinholdMetricGeometryEquilibrium1975b}%
  \BibitemOpen
  \bibfield  {author} {\bibinfo {author} {\bibfnamefont {F.}~\bibnamefont
  {Weinhold}},\ }\href {\doibase 10.1063/1.431636} {\bibfield  {journal}
  {\bibinfo  {journal} {J. Chem. Phys.}\ }\textbf {\bibinfo {volume} {63}},\
  \bibinfo {pages} {2488} (\bibinfo {year} {1975}{\natexlab{b}})}\BibitemShut
  {NoStop}%
\bibitem [{\citenamefont {Ruppeiner}(1979)}]{Ruppeiner1979}%
  \BibitemOpen
  \bibfield  {author} {\bibinfo {author} {\bibfnamefont {G.}~\bibnamefont
  {Ruppeiner}},\ }\href {\doibase 10.1103/PhysRevA.20.1608} {\bibfield
  {journal} {\bibinfo  {journal} {Phys. Rev. A}\ }\textbf {\bibinfo {volume}
  {20}},\ \bibinfo {pages} {1608} (\bibinfo {year} {1979})}\BibitemShut
  {NoStop}%
\bibitem [{\citenamefont {Salamon}\ and\ \citenamefont
  {Berry}(1983)}]{Salamon1983a}%
  \BibitemOpen
  \bibfield  {author} {\bibinfo {author} {\bibfnamefont {P.}~\bibnamefont
  {Salamon}}\ and\ \bibinfo {author} {\bibfnamefont {R.~S.}\ \bibnamefont
  {Berry}},\ }\href@noop {} {\bibfield  {journal} {\bibinfo  {journal} {Phys.
  Rev. Lett.}\ }\textbf {\bibinfo {volume} {51}},\ \bibinfo {pages} {1127}
  (\bibinfo {year} {1983})}\BibitemShut {NoStop}%
\bibitem [{\citenamefont {Schl{\"o}gl}(1985)}]{Schlogl1985}%
  \BibitemOpen
  \bibfield  {author} {\bibinfo {author} {\bibfnamefont {F.}~\bibnamefont
  {Schl{\"o}gl}},\ }\href {\doibase 10.1007/BF01328857} {\bibfield  {journal}
  {\bibinfo  {journal} {Zeitschrift f{\"u}r Physik B}\ }\textbf {\bibinfo
  {volume} {59}},\ \bibinfo {pages} {449} (\bibinfo {year} {1985})}\BibitemShut
  {NoStop}%
\bibitem [{\citenamefont {Ruppeiner}(1995)}]{Ruppeiner}%
  \BibitemOpen
  \bibfield  {author} {\bibinfo {author} {\bibfnamefont {G.}~\bibnamefont
  {Ruppeiner}},\ }\href@noop {} {\bibfield  {journal} {\bibinfo  {journal}
  {Rev. Mod. Phys.}\ }\textbf {\bibinfo {volume} {67}},\ \bibinfo {pages} {605}
  (\bibinfo {year} {1995})}\BibitemShut {NoStop}%
\bibitem [{\citenamefont {Crooks}(2007)}]{Crooks2007}%
  \BibitemOpen
  \bibfield  {author} {\bibinfo {author} {\bibfnamefont {G.~E.}\ \bibnamefont
  {Crooks}},\ }\href {\doibase 10.1103/PhysRevLett.99.100602} {\bibfield
  {journal} {\bibinfo  {journal} {Phys. Rev. Lett.}\ }\textbf {\bibinfo
  {volume} {99}},\ \bibinfo {pages} {100602} (\bibinfo {year}
  {2007})}\BibitemShut {NoStop}%
\bibitem [{\citenamefont {Sivak}\ and\ \citenamefont
  {Crooks}(2012)}]{Sivak2012a}%
  \BibitemOpen
  \bibfield  {author} {\bibinfo {author} {\bibfnamefont {D.~A.}\ \bibnamefont
  {Sivak}}\ and\ \bibinfo {author} {\bibfnamefont {G.~E.}\ \bibnamefont
  {Crooks}},\ }\href {\doibase 10.1103/PhysRevLett.108.190602} {\bibfield
  {journal} {\bibinfo  {journal} {Phys. Rev. L}\ }\textbf {\bibinfo {volume}
  {108}},\ \bibinfo {pages} {190602 (2012)} (\bibinfo {year}
  {2012})}\BibitemShut {NoStop}%
\bibitem [{\citenamefont {Zulkowski}\ \emph {et~al.}(2012)\citenamefont
  {Zulkowski}, \citenamefont {Sivak}, \citenamefont {Crooks},\ and\
  \citenamefont {Deweese}}]{Zulkowski2012}%
  \BibitemOpen
  \bibfield  {author} {\bibinfo {author} {\bibfnamefont {P.~R.}\ \bibnamefont
  {Zulkowski}}, \bibinfo {author} {\bibfnamefont {D.~A.}\ \bibnamefont
  {Sivak}}, \bibinfo {author} {\bibfnamefont {G.~E.}\ \bibnamefont {Crooks}}, \
  and\ \bibinfo {author} {\bibfnamefont {M.~R.}\ \bibnamefont {Deweese}},\
  }\href {\doibase 10.1103/PhysRevE.86.041148} {\bibfield  {journal} {\bibinfo
  {journal} {Phys. Rev. E}\ }\textbf {\bibinfo {volume} {86}},\ \bibinfo
  {pages} {0141148} (\bibinfo {year} {2012})}\BibitemShut {NoStop}%
\bibitem [{\citenamefont {Scandi}\ and\ \citenamefont
  {Perarnau-Llobet}(2019)}]{Scandi}%
  \BibitemOpen
  \bibfield  {author} {\bibinfo {author} {\bibfnamefont {M.}~\bibnamefont
  {Scandi}}\ and\ \bibinfo {author} {\bibfnamefont {M.}~\bibnamefont
  {Perarnau-Llobet}},\ }\href@noop {} {\bibfield  {journal} {\bibinfo
  {journal} {Quantum}\ }\textbf {\bibinfo {volume} {3}},\ \bibinfo {pages}
  {197} (\bibinfo {year} {2019})}\BibitemShut {NoStop}%
\bibitem [{\citenamefont {Deffner}\ and\ \citenamefont
  {Bonanca}(2020)}]{Deffner}%
  \BibitemOpen
  \bibfield  {author} {\bibinfo {author} {\bibfnamefont {S.}~\bibnamefont
  {Deffner}}\ and\ \bibinfo {author} {\bibfnamefont {M.~V.~S.}\ \bibnamefont
  {Bonanca}},\ }\href@noop {} {\bibfield  {journal} {\bibinfo  {journal} {EPL}\
  }\textbf {\bibinfo {volume} {131}},\ \bibinfo {pages} {20001} (\bibinfo
  {year} {2020})}\BibitemShut {NoStop}%
\bibitem [{\citenamefont {Miller}\ \emph {et~al.}(2019)\citenamefont {Miller},
  \citenamefont {Scandi}, \citenamefont {Anders},\ and\ \citenamefont
  {Perarnau-Llobet}}]{Miller2019}%
  \BibitemOpen
  \bibfield  {author} {\bibinfo {author} {\bibfnamefont {H.~J.~D.}\
  \bibnamefont {Miller}}, \bibinfo {author} {\bibfnamefont {M.}~\bibnamefont
  {Scandi}}, \bibinfo {author} {\bibfnamefont {J.}~\bibnamefont {Anders}}, \
  and\ \bibinfo {author} {\bibfnamefont {M.}~\bibnamefont {Perarnau-Llobet}},\
  }\href@noop {} {\bibfield  {journal} {\bibinfo  {journal} {Phys. Rev. Lett.}\
  }\textbf {\bibinfo {volume} {123}},\ \bibinfo {pages} {230603} (\bibinfo
  {year} {2019})}\BibitemShut {NoStop}%
\bibitem [{\citenamefont {Miettinen}(1999)}]{Miettinen1999}%
  \BibitemOpen
  \bibfield  {author} {\bibinfo {author} {\bibfnamefont {K.}~\bibnamefont
  {Miettinen}},\ }\href@noop {} {\emph {\bibinfo {title} {{Nonlinear
  Multiobjective Optimization}}}}\ (\bibinfo  {publisher} {Springer Science
  {\&} Business Media},\ \bibinfo {year} {1999})\BibitemShut {NoStop}%
\bibitem [{\citenamefont {Jarzynski}(1997)}]{Jarzynski1997d}%
  \BibitemOpen
  \bibfield  {author} {\bibinfo {author} {\bibfnamefont {C.}~\bibnamefont
  {Jarzynski}},\ }\href {\doibase 10.1103/PhysRevLett.78.2690} {\bibfield
  {journal} {\bibinfo  {journal} {Phys. Rev. Lett.}\ }\textbf {\bibinfo
  {volume} {78}},\ \bibinfo {pages} {2690} (\bibinfo {year}
  {1997})}\BibitemShut {NoStop}%
\bibitem [{\citenamefont {Speck}\ and\ \citenamefont {Seifert}(2004)}]{Speck}%
  \BibitemOpen
  \bibfield  {author} {\bibinfo {author} {\bibfnamefont {T.}~\bibnamefont
  {Speck}}\ and\ \bibinfo {author} {\bibfnamefont {U.}~\bibnamefont
  {Seifert}},\ }\href@noop {} {\bibfield  {journal} {\bibinfo  {journal} {Phys.
  Rev. E}\ }\textbf {\bibinfo {volume} {70}},\ \bibinfo {pages} {066112}
  (\bibinfo {year} {2004})}\BibitemShut {NoStop}%
\bibitem [{\citenamefont {Mandal}\ and\ \citenamefont
  {Jarzynski}(2016)}]{Mandal2016a}%
  \BibitemOpen
  \bibfield  {author} {\bibinfo {author} {\bibfnamefont {D.}~\bibnamefont
  {Mandal}}\ and\ \bibinfo {author} {\bibfnamefont {C.}~\bibnamefont
  {Jarzynski}},\ }\href {\doibase 10.1088/1742-5468/2016/06/063204} {\bibfield
  {journal} {\bibinfo  {journal} {J. Stat. Mech.}\ }\textbf {\bibinfo {volume}
  {2016}},\ \bibinfo {pages} {063204} (\bibinfo {year} {2016})}\BibitemShut
  {NoStop}%
\bibitem [{\citenamefont {Nulton}\ \emph {et~al.}(1985)\citenamefont {Nulton},
  \citenamefont {Salamon}, \citenamefont {Andresen}, \citenamefont {Anmin},
  \citenamefont {Nulton}, \citenamefont {Salamon}, \citenamefont {Andresen},\
  and\ \citenamefont {Anmin}}]{Nulton2013}%
  \BibitemOpen
  \bibfield  {author} {\bibinfo {author} {\bibfnamefont {J.}~\bibnamefont
  {Nulton}}, \bibinfo {author} {\bibfnamefont {P.}~\bibnamefont {Salamon}},
  \bibinfo {author} {\bibfnamefont {B.}~\bibnamefont {Andresen}}, \bibinfo
  {author} {\bibfnamefont {Q.}~\bibnamefont {Anmin}}, \bibinfo {author}
  {\bibfnamefont {J.}~\bibnamefont {Nulton}}, \bibinfo {author} {\bibfnamefont
  {P.}~\bibnamefont {Salamon}}, \bibinfo {author} {\bibfnamefont
  {B.}~\bibnamefont {Andresen}}, \ and\ \bibinfo {author} {\bibfnamefont
  {Q.}~\bibnamefont {Anmin}},\ }\href {\doibase 10.1063/1.449774} {\bibfield
  {journal} {\bibinfo  {journal} {J. Chem. Phys.}\ }\textbf {\bibinfo {volume}
  {83}},\ \bibinfo {pages} {334} (\bibinfo {year} {1985})}\BibitemShut
  {NoStop}%
\bibitem [{\citenamefont {Anders}\ and\ \citenamefont
  {Giovannetti}(2013)}]{Anders2013a}%
  \BibitemOpen
  \bibfield  {author} {\bibinfo {author} {\bibfnamefont {J.}~\bibnamefont
  {Anders}}\ and\ \bibinfo {author} {\bibfnamefont {V.}~\bibnamefont
  {Giovannetti}},\ }\href {\doibase 10.1088/1367-2630/15/3/033022} {\bibfield
  {journal} {\bibinfo  {journal} {N. J. Phys}\ }\textbf {\bibinfo {volume}
  {15}},\ \bibinfo {pages} {033022} (\bibinfo {year} {2013})}\BibitemShut
  {NoStop}%
\bibitem [{\citenamefont {Large}\ and\ \citenamefont
  {Sivak}(2019)}]{Large2019}%
  \BibitemOpen
  \bibfield  {author} {\bibinfo {author} {\bibfnamefont {S.~J.}\ \bibnamefont
  {Large}}\ and\ \bibinfo {author} {\bibfnamefont {D.~A.}\ \bibnamefont
  {Sivak}},\ }\href@noop {} {\bibfield  {journal} {\bibinfo  {journal} {J.
  Stat. Mech.}\ }\textbf {\bibinfo {volume} {2019}},\ \bibinfo {pages} {083212}
  (\bibinfo {year} {2019})}\BibitemShut {NoStop}%
\bibitem [{\citenamefont {Scandi}\ \emph {et~al.}(2020)\citenamefont {Scandi},
  \citenamefont {Miller}, \citenamefont {Anders},\ and\ \citenamefont
  {Perarnau-Llobet}}]{Scandi2019}%
  \BibitemOpen
  \bibfield  {author} {\bibinfo {author} {\bibfnamefont {M.}~\bibnamefont
  {Scandi}}, \bibinfo {author} {\bibfnamefont {H.~J.~D.}\ \bibnamefont
  {Miller}}, \bibinfo {author} {\bibfnamefont {J.}~\bibnamefont {Anders}}, \
  and\ \bibinfo {author} {\bibfnamefont {M.}~\bibnamefont {Perarnau-Llobet}},\
  }\href@noop {} {\bibfield  {journal} {\bibinfo  {journal} {Phys. Rev.
  Research}\ }\textbf {\bibinfo {volume} {2}},\ \bibinfo {pages} {023377}
  (\bibinfo {year} {2020})}\BibitemShut {NoStop}%
\bibitem [{\citenamefont {Weedbrook}\ \emph {et~al.}(2012)\citenamefont
  {Weedbrook}, \citenamefont {Pirandola}, \citenamefont {Cerf},\ and\
  \citenamefont {Ralph}}]{Weedbrook}%
  \BibitemOpen
  \bibfield  {author} {\bibinfo {author} {\bibfnamefont {C.}~\bibnamefont
  {Weedbrook}}, \bibinfo {author} {\bibfnamefont {S.}~\bibnamefont
  {Pirandola}}, \bibinfo {author} {\bibfnamefont {N.~J.}\ \bibnamefont {Cerf}},
  \ and\ \bibinfo {author} {\bibfnamefont {T.~C.}\ \bibnamefont {Ralph}},\
  }\href@noop {} {\bibfield  {journal} {\bibinfo  {journal} {Rev. Mod. Phys.}\
  }\textbf {\bibinfo {volume} {84}},\ \bibinfo {pages} {621} (\bibinfo {year}
  {2012})}\BibitemShut {NoStop}%
\bibitem [{\citenamefont {Brandner}\ and\ \citenamefont
  {Seifert}(2016)}]{Brandner2016a}%
  \BibitemOpen
  \bibfield  {author} {\bibinfo {author} {\bibfnamefont {K.}~\bibnamefont
  {Brandner}}\ and\ \bibinfo {author} {\bibfnamefont {U.}~\bibnamefont
  {Seifert}},\ }\href@noop {} {\bibfield  {journal} {\bibinfo  {journal} {Phys.
  Rev. E}\ }\textbf {\bibinfo {volume} {93}},\ \bibinfo {pages} {062134}
  (\bibinfo {year} {2016})}\BibitemShut {NoStop}%
\bibitem [{\citenamefont {Brandner}\ \emph {et~al.}(2015)\citenamefont
  {Brandner}, \citenamefont {Saito},\ and\ \citenamefont
  {Seifert}}]{Brandner2018}%
  \BibitemOpen
  \bibfield  {author} {\bibinfo {author} {\bibfnamefont {K.}~\bibnamefont
  {Brandner}}, \bibinfo {author} {\bibfnamefont {K.}~\bibnamefont {Saito}}, \
  and\ \bibinfo {author} {\bibfnamefont {U.}~\bibnamefont {Seifert}},\
  }\href@noop {} {\bibfield  {journal} {\bibinfo  {journal} {Phys. Rev. X}\
  }\textbf {\bibinfo {volume} {5}},\ \bibinfo {pages} {031019} (\bibinfo {year}
  {2015})}\BibitemShut {NoStop}%
\bibitem [{\citenamefont {Talkner}\ \emph {et~al.}(2007)\citenamefont
  {Talkner}, \citenamefont {Lutz},\ and\ \citenamefont
  {H{\"{a}}nggi}}]{Talkner2007c}%
  \BibitemOpen
  \bibfield  {author} {\bibinfo {author} {\bibfnamefont {P.}~\bibnamefont
  {Talkner}}, \bibinfo {author} {\bibfnamefont {E.}~\bibnamefont {Lutz}}, \
  and\ \bibinfo {author} {\bibfnamefont {P.}~\bibnamefont {H{\"{a}}nggi}},\
  }\href {\doibase 10.1103/PhysRevE.75.050102} {\bibfield  {journal} {\bibinfo
  {journal} {Phys. Rev. E}\ }\textbf {\bibinfo {volume} {75}},\ \bibinfo
  {pages} {050102} (\bibinfo {year} {2007})}\BibitemShut {NoStop}%
\bibitem [{\citenamefont {Petz}(1994)}]{Petz1994a}%
  \BibitemOpen
  \bibfield  {author} {\bibinfo {author} {\bibfnamefont {D.}~\bibnamefont
  {Petz}},\ }\href {\doibase 10.1063/1.530611} {\bibfield  {journal} {\bibinfo
  {journal} {J. Math. Phys.}\ }\textbf {\bibinfo {volume} {35}},\ \bibinfo
  {pages} {780} (\bibinfo {year} {1994})}\BibitemShut {NoStop}%
\bibitem [{\citenamefont {Hayashi}(2002)}]{Hayashi2002}%
  \BibitemOpen
  \bibfield  {author} {\bibinfo {author} {\bibfnamefont {M.}~\bibnamefont
  {Hayashi}},\ }\href@noop {} {\bibfield  {journal} {\bibinfo  {journal} {J.
  Phys. A}\ }\textbf {\bibinfo {volume} {35}},\ \bibinfo {pages} {7689}
  (\bibinfo {year} {2002})}\BibitemShut {NoStop}%
\bibitem [{Note1()}]{Note1}%
  \BibitemOpen
  \bibinfo {note} {In this case one has $t\propto s(\phi _t^\epsilon )$, where
  $s(\phi _t^\epsilon )$ is the arc length for the interval $[0,\phi
  _t^\epsilon ]$. This means the curve $\gamma :t\DOTSB \mapstochar \rightarrow
  \protect \mathaccentV {vec}17E{\Lambda }_t$ is traversed at constant
  velocity, leading to the equality condition for the Cauchy-Schwartz
  inequality.}\BibitemShut {Stop}%
\bibitem [{\citenamefont {Ro{\ss}nagel}\ \emph {et~al.}(2015)\citenamefont
  {Ro{\ss}nagel}, \citenamefont {Dawkins}, \citenamefont {Tolazzi},
  \citenamefont {Abah}, \citenamefont {Lutz}, \citenamefont {Schmidt-Kaler},\
  and\ \citenamefont {Singer}}]{Rossnagel2015}%
  \BibitemOpen
  \bibfield  {author} {\bibinfo {author} {\bibfnamefont {J.}~\bibnamefont
  {Ro{\ss}nagel}}, \bibinfo {author} {\bibfnamefont {S.~T.}\ \bibnamefont
  {Dawkins}}, \bibinfo {author} {\bibfnamefont {K.~N.}\ \bibnamefont
  {Tolazzi}}, \bibinfo {author} {\bibfnamefont {O.}~\bibnamefont {Abah}},
  \bibinfo {author} {\bibfnamefont {E.}~\bibnamefont {Lutz}}, \bibinfo {author}
  {\bibfnamefont {F.}~\bibnamefont {Schmidt-Kaler}}, \ and\ \bibinfo {author}
  {\bibfnamefont {K.}~\bibnamefont {Singer}},\ }\href@noop {} {\bibfield
  {journal} {\bibinfo  {journal} {Science}\ }\textbf {\bibinfo {volume}
  {352}},\ \bibinfo {pages} {325} (\bibinfo {year} {2015})}\BibitemShut
  {NoStop}%
\bibitem [{\citenamefont {Mehboudi}\ and\ \citenamefont
  {Parrondo}(2019)}]{Mehboudi}%
  \BibitemOpen
  \bibfield  {author} {\bibinfo {author} {\bibfnamefont {M.}~\bibnamefont
  {Mehboudi}}\ and\ \bibinfo {author} {\bibfnamefont {J.~M.~R.}\ \bibnamefont
  {Parrondo}},\ }\href@noop {} {\bibfield  {journal} {\bibinfo  {journal} {N.
  J. Phys}\ }\textbf {\bibinfo {volume} {21}},\ \bibinfo {pages} {083036}
  (\bibinfo {year} {2019})}\BibitemShut {NoStop}%
\bibitem [{Note2()}]{Note2}%
  \BibitemOpen
  \bibinfo {note} {The metric has a long and cumbersome analytical expression
  that we do not present.}\BibitemShut {Stop}%
\bibitem [{Note3()}]{Note3}%
  \BibitemOpen
  \bibinfo {note} {For a convex Pareto front, the full set of minima of
  $\protect \mathcal {I}^*_\epsilon \ \ \forall \epsilon \in [0,1]$
  characterise the entire front. However, points situated along any non-convex
  regions of the front cannot be determined from minimising $\protect \mathcal
  {I}_\epsilon $ alone. In such situations, one can additionally apply an
  adaptive weighted sum method \cite {Kim2005} to determine any non-convex
  sections of the front.}\BibitemShut {Stop}%
\bibitem [{\citenamefont {Newman}\ \emph {et~al.}(2017)\citenamefont {Newman},
  \citenamefont {Mintert},\ and\ \citenamefont {Nazir}}]{Newman2017}%
  \BibitemOpen
  \bibfield  {author} {\bibinfo {author} {\bibfnamefont {D.}~\bibnamefont
  {Newman}}, \bibinfo {author} {\bibfnamefont {F.}~\bibnamefont {Mintert}}, \
  and\ \bibinfo {author} {\bibfnamefont {A.}~\bibnamefont {Nazir}},\
  }\href@noop {} {\bibfield  {journal} {\bibinfo  {journal} {Phys. Rev. E}\
  }\textbf {\bibinfo {volume} {95}},\ \bibinfo {pages} {032139} (\bibinfo
  {year} {2017})}\BibitemShut {NoStop}%
\bibitem [{\citenamefont {Deffner}\ and\ \citenamefont
  {Lutz}(2018)}]{Deffner2018b}%
  \BibitemOpen
  \bibfield  {author} {\bibinfo {author} {\bibfnamefont {S.}~\bibnamefont
  {Deffner}}\ and\ \bibinfo {author} {\bibfnamefont {E.}~\bibnamefont {Lutz}},\
  }\href@noop {} {\bibfield  {journal} {\bibinfo  {journal} {Phys. Rev. E}\
  }\textbf {\bibinfo {volume} {87}},\ \bibinfo {pages} {022143} (\bibinfo
  {year} {2018})}\BibitemShut {NoStop}%
\bibitem [{\citenamefont {Kim}\ and\ \citenamefont {Weck}(2005)}]{Kim2005}%
  \BibitemOpen
  \bibfield  {author} {\bibinfo {author} {\bibfnamefont {I.~Y.}\ \bibnamefont
  {Kim}}\ and\ \bibinfo {author} {\bibfnamefont {O.~L.~D.}\ \bibnamefont
  {Weck}},\ }\href {\doibase 10.1007/s00158-004-0465-1} {\ \textbf {\bibinfo
  {volume} {158}},\ \bibinfo {pages} {149} (\bibinfo {year}
  {2005})}\BibitemShut {NoStop}%
\bibitem [{\citenamefont {Tomamichel}\ \emph {et~al.}(2013)\citenamefont
  {Tomamichel}, \citenamefont {Hayashi},\ and\ \citenamefont
  {Member}}]{Tomamichel2013a}%
  \BibitemOpen
  \bibfield  {author} {\bibinfo {author} {\bibfnamefont {M.}~\bibnamefont
  {Tomamichel}}, \bibinfo {author} {\bibfnamefont {M.}~\bibnamefont {Hayashi}},
  \ and\ \bibinfo {author} {\bibfnamefont {S.}~\bibnamefont {Member}},\ }\href
  {\doibase 10.1109/TIT.2013.2276628} {\bibfield  {journal} {\bibinfo
  {journal} {IEEE Trans. Inform. Theory}\ }\textbf {\bibinfo {volume} {59}},\
  \bibinfo {pages} {7693} (\bibinfo {year} {2013})}\BibitemShut {NoStop}%
\bibitem [{\citenamefont {Janyszek}(1986)}]{Unicersity}%
  \BibitemOpen
  \bibfield  {author} {\bibinfo {author} {\bibfnamefont {H.}~\bibnamefont
  {Janyszek}},\ }\href@noop {} {\bibfield  {journal} {\bibinfo  {journal} {Rep.
  Math. Phys.}\ }\textbf {\bibinfo {volume} {24}},\ \bibinfo {pages} {11}
  (\bibinfo {year} {1986})}\BibitemShut {NoStop}%
\bibitem [{\citenamefont {Alicki}\ and\ \citenamefont
  {Lendi}(2007)}]{Holevo2001}%
  \BibitemOpen
  \bibfield  {author} {\bibinfo {author} {\bibfnamefont {R.}~\bibnamefont
  {Alicki}}\ and\ \bibinfo {author} {\bibfnamefont {K.}~\bibnamefont {Lendi}},\
  }\href@noop {} {\emph {\bibinfo {title} {{Quantum Dynamical Semigroups and
  Applications}}}}\ (\bibinfo  {publisher} {Springer},\ \bibinfo {year}
  {2007})\BibitemShut {NoStop}%
\bibitem [{\citenamefont {Cavina}\ \emph {et~al.}(2017)\citenamefont {Cavina},
  \citenamefont {Mari},\ and\ \citenamefont {Giovannetti}}]{Cavina2017}%
  \BibitemOpen
  \bibfield  {author} {\bibinfo {author} {\bibfnamefont {V.}~\bibnamefont
  {Cavina}}, \bibinfo {author} {\bibfnamefont {A.}~\bibnamefont {Mari}}, \ and\
  \bibinfo {author} {\bibfnamefont {V.}~\bibnamefont {Giovannetti}},\ }\href
  {\doibase 10.1103/PhysRevLett.119.050601} {\bibfield  {journal} {\bibinfo
  {journal} {Phys. Rev. Lett.}\ }\textbf {\bibinfo {volume} {119}},\ \bibinfo
  {pages} {050601} (\bibinfo {year} {2017})}\BibitemShut {NoStop}%
\bibitem [{\citenamefont {Miller}\ \emph {et~al.}()\citenamefont {Miller},
  \citenamefont {Mohammady}, \citenamefont {Perarnau-Llobet},\ and\
  \citenamefont {Guarnieri}}]{Miller}%
  \BibitemOpen
  \bibfield  {author} {\bibinfo {author} {\bibfnamefont {H.~J.~D.}\
  \bibnamefont {Miller}}, \bibinfo {author} {\bibfnamefont {M.~H.}\
  \bibnamefont {Mohammady}}, \bibinfo {author} {\bibfnamefont {M.}~\bibnamefont
  {Perarnau-Llobet}}, \ and\ \bibinfo {author} {\bibfnamefont {G.}~\bibnamefont
  {Guarnieri}},\ }\href@noop {} {\ }\Eprint
  {http://arxiv.org/abs/arXiv:2006.07316v2} {arXiv:arXiv:2006.07316v2}
  \BibitemShut {NoStop}%
\bibitem [{\citenamefont {Leggio}\ \emph {et~al.}(2013)\citenamefont {Leggio},
  \citenamefont {Napoli}, \citenamefont {Messina},\ and\ \citenamefont
  {Breuer}}]{Leggio2013a}%
  \BibitemOpen
  \bibfield  {author} {\bibinfo {author} {\bibfnamefont {B.}~\bibnamefont
  {Leggio}}, \bibinfo {author} {\bibfnamefont {A.}~\bibnamefont {Napoli}},
  \bibinfo {author} {\bibfnamefont {A.}~\bibnamefont {Messina}}, \ and\
  \bibinfo {author} {\bibfnamefont {H.-P.}\ \bibnamefont {Breuer}},\ }\href
  {\doibase 10.1103/PhysRevA.88.042111} {\bibfield  {journal} {\bibinfo
  {journal} {Phys. Rev. A}\ }\textbf {\bibinfo {volume} {88}},\ \bibinfo
  {pages} {042111} (\bibinfo {year} {2013})}\BibitemShut {NoStop}%
\bibitem [{\citenamefont {Horowitz}\ and\ \citenamefont
  {Parrondo}(2013)}]{Horowitz2013b}%
  \BibitemOpen
  \bibfield  {author} {\bibinfo {author} {\bibfnamefont {J.~M.}\ \bibnamefont
  {Horowitz}}\ and\ \bibinfo {author} {\bibfnamefont {J.~M.~R.}\ \bibnamefont
  {Parrondo}},\ }\href {\doibase 10.1088/1367-2630/15/8/085028} {\bibfield
  {journal} {\bibinfo  {journal} {N. J. Phys}\ }\textbf {\bibinfo {volume}
  {15}},\ \bibinfo {pages} {085028} (\bibinfo {year} {2013})}\BibitemShut
  {NoStop}%
\bibitem [{\citenamefont {Manzano}\ \emph {et~al.}(2015)\citenamefont
  {Manzano}, \citenamefont {Horowitz},\ and\ \citenamefont
  {Parrondo}}]{Manzano2015}%
  \BibitemOpen
  \bibfield  {author} {\bibinfo {author} {\bibfnamefont {G.}~\bibnamefont
  {Manzano}}, \bibinfo {author} {\bibfnamefont {J.~M.}\ \bibnamefont
  {Horowitz}}, \ and\ \bibinfo {author} {\bibfnamefont {J.~M.~R.}\ \bibnamefont
  {Parrondo}},\ }\href {\doibase 10.1103/PhysRevE.92.032129} {\bibfield
  {journal} {\bibinfo  {journal} {Phys. Rev. E}\ }\textbf {\bibinfo {volume}
  {92}},\ \bibinfo {pages} {032129} (\bibinfo {year} {2015})}\BibitemShut
  {NoStop}%
\end{thebibliography}%

\newpage

\widetext

\appendix

\section{Thermodynamic metrics and Fisher information}

\

In this Appendix we provide further background concerning the information-geometric interpretations of the metric tensors (9) and (11) for both classical and quantum mechanical systems.

Consider first a statistical manifold parameterised by coordinates $\vec{\Lambda}$, with each coordinate defining a normalised probability distribution $p(\vec{\Lambda})$ with outcome space $\chi$. We denote the expectation value of an observable $A$ with respect to $p(\vec{\Lambda})$ as $\langle A \rangle=\int dz \  p(z|\vec{\Lambda})A(z)$, where $z\in\chi$ are points in the outcome space. A particular divergence measure between two distributions on the manifold is the Kullback-Liebler divergence:
\begin{align}
	D[p(\vec{\Lambda})|| p(\vec{\Lambda}')]:=\big< \text{ln}  \ p(\vec{\Lambda})/p(\vec{\Lambda}')  \big>\geq 0,
\end{align}
which quantifies how distinguishable two distributions are from each other. If one considers two close points $\vec{\Lambda}$ and $\vec{\Lambda}+d\vec{\Lambda}$, the divergence between the distributions becomes
\begin{align}\label{eq:KL}
	D[p(\vec{\Lambda})|| p(\vec{\Lambda}+d\vec{\Lambda})]=\frac{1}{2} F_{jk}(\vec{\Lambda}) d\Lambda^j d\Lambda^k+\mathcal{O}(|d\vec{\Lambda}|^3),
\end{align}
where
\begin{align}\label{eq:fisher}
	F_{jk}(\vec{\Lambda}):=\bigg< \frac{\partial}{\partial \Lambda_j} \text{ln} \ p(\vec{\Lambda})  \frac{\partial}{\partial \Lambda_k} \text{ln} \ p(\vec{\Lambda}) \bigg>,
\end{align}
is known as the \textit{Fisher-Rao matrix}. We may interpret this as a Riemann metric tensor on the manifold, since it is positive semi-definite, symmetric and smooth in $\vec{\Lambda}$. This implies a notion of distance between points in the manifold, with squared line element $ds^2=F_{jk}(\vec{\Lambda})d\Lambda^j d\Lambda^k$.

For quantum systems, one may instead consider a manifold of normalised density matrices $\rho(\vec{\Lambda})$. The quantum relative entropy replaces~\eqref{eq:KL}, defined as
\begin{align}
	S[\rho(\vec{\Lambda})|| \rho(\vec{\Lambda}')]:=\tr{\rho (\vec{\Lambda}) \text{ln} \ \rho(\vec{\Lambda})}-\tr{\rho (\vec{\Lambda}) \text{ln} \ \rho(\vec{\Lambda}')}.
\end{align}
Considering again close points $\vec{\Lambda}$ and $\vec{\Lambda}+d\vec{\Lambda}$, we have
\begin{align}\label{eq:rel_expand}
	S[\rho(\vec{\Lambda})|| \rho(\vec{\Lambda}+d\vec{\Lambda})]=\frac{1}{2} \mathcal{F}_{jk}(\vec{\Lambda}) d\Lambda^j d\Lambda^k+\mathcal{O}(|d\vec{\Lambda}|^3),
\end{align}
where
\begin{align}\label{eq:kubo}
	\mathcal{F}_{jk}(\vec{\Lambda}):=\int^1_0 ds \ \text{Tr}\bigg(\bigg[\frac{\partial}{\partial \Lambda_j} \text{ln} \ \rho(\vec{\Lambda})\bigg] \rho^s(\vec{\Lambda})\bigg[\frac{\partial}{\partial \Lambda_k} \text{ln} \ \rho(\vec{\Lambda})\bigg]\rho^{1-s}(\vec{\Lambda})\bigg),
\end{align}
is known as the Kubo-Mori Fisher information matrix \cite{Petz1994a,Hayashi2002}. This provides a measure of distance between neighbouring density matrices on the manifold with squared line element $ds^2=\mathcal{F}_{jk}(\vec{\Lambda})d\Lambda^j d\Lambda^k$. We note however, that other information metrics may be obtained by replacing the relative entropy by a different choice of divergence. For example, consider instead the \textit{relative entropy variance} \cite{Tomamichel2013a}
\begin{align}
	V[\rho(\vec{\Lambda})|| \rho(\vec{\Lambda}')]:=\text{Tr}\bigg(\rho (\vec{\Lambda}) \big(\text{ln} \ \rho(\vec{\Lambda})-\text{ln} \ \rho(\vec{\Lambda}')\big)^2\bigg)-S^2[\rho(\vec{\Lambda})|| \rho(\vec{\Lambda}')].
\end{align}
Expanding this between neighbouring states gives \begin{align}
	V[\rho(\vec{\Lambda})|| \rho(\vec{\Lambda}+d\vec{\Lambda})]= \tilde{\mathcal{F}}_{jk}(\vec{\Lambda}) d\Lambda^j d\Lambda^k+\mathcal{O}(|d\vec{\Lambda}|^3),
\end{align}
where
\begin{align}\label{eq:relvar}
	\tilde{\mathcal{F}}_{jk}(\vec{\Lambda}):=\frac{1}{2}\text{Tr}\bigg(\bigg\{\frac{\partial}{\partial \Lambda_j} \text{ln} \ \rho(\vec{\Lambda}) ,\frac{\partial}{\partial \Lambda_k} \text{ln} \ \rho(\vec{\Lambda})\bigg\}\rho(\vec{\Lambda})\bigg)
\end{align}
defines another metric. This choice was first introduced in \cite{Unicersity} as an alternative to the Kubo-Mori metric. Note that for quasi-classical states with a spectral decomposition of the form $\rho(\vec{\Lambda})=\sum_n p_n(\vec{\Lambda})\ket{n}\bra{n}$, with eigenstates $\{\ket{n}\}$ that are independent of coordinates $\vec{\Lambda}$, both metrics~\eqref{eq:kubo} and~\eqref{eq:relvar} become proportional to the Fisher-Rao metric~\eqref{eq:fisher}. The difference between these two metrics highlights the role of quantum coherence. If we restrict our attention to the manifold of thermal states $\rho(\vec{\Lambda})=\pi(\vec{\Lambda}):=\exp{(-\beta H(\vec{\lambda}))}/\tr{\exp{(-\beta H(\vec{\lambda}))}}$ and introduce conjugate forces $\{X_i \}$ defined in (2), we see this difference between the metrics occurs when at least one pair of forces are non-commuting, $[X_i,X_j]\neq 0$. By comparing the different metric expressions used in the main text, we find that the work fluctuations (9) are determined by components of the metric~\eqref{eq:relvar}, while the efficiency (11) is determined by the Kubo-Mori metric~\eqref{eq:kubo}.    

\section{Derivation of (8) and (10)}

\

In this section we provide details of the derivations of the metric expression for work fluctuations and efficiency. Let us begin by considering the protocol $\gamma:t\mapsto \vec{\Lambda}_t$ with $t\in[0,1]$, evaluated at $N$ discrete points $t_n=(n-1)/(N-1)$. We introduce the operator $\Delta \Phi_n/N=\Phi(\vec{\Lambda}_{t_{n+1}})-\Phi(\vec{\Lambda}_{t_{n}})$, where $\Phi(\vec{\Lambda})=\beta H(\vec{\lambda})+\text{ln} \mathcal{Z}(\vec{\Lambda})$ is the non-equilibrium potential with $\text{ln} \mathcal{Z}(\vec{\Lambda})=\tr{\exp{(-\beta H(\vec{\lambda}))}}$ the partition function. Similarly we define $\Delta H_n/N=H(\vec{\lambda}_{t_{n+1}})-H(\vec{\lambda}_{t_{n}})$. It is then straightforward to see that the work fluctuations (7) to leading order in $1/N$ are
\begin{align}\label{eq:var_expand}
	\nonumber\text{Var}(W)&=\frac{1}{N}\lim_{N\to\infty}\sum^{N-1}_{n=1} \bigg(\frac{1}{N}\bigg)\tr{\Delta H_n^2 \pi(\vec{\Lambda}_{t_n})}-\tr{\Delta H_n \pi(\vec{\Lambda}_{t_n})}^2+\mathcal{O}(1/N^2), \\
	&=\frac{1}{N}\int^1_0 dt \ \tr{\dot{H}^2(\vec{\lambda}_t)\pi(\vec{\Lambda})}-\tr{\dot{H}(\vec{\lambda}_t)\pi(\vec{\Lambda})}^2+\mathcal{O}(1/N^2),
\end{align}
where we denote $\dot{H}(\vec{\lambda})=(\partial/\partial t) H(\vec{\lambda})$. Expanding $\dot{H}(\vec{\lambda})=\dot{\lambda}^j X_j$ completes the derivation of the metric (8). For the efficiency, we first derive the expression for irreversible entropy production (3),
\begin{align}
	\nonumber S_{irr}&=\sum^{N-1}_{n=1}S\big(\pi(\vec{\Lambda}_{t_{n}})||\pi(\vec{\Lambda}_{t_{n+1}})\big), \\
	\nonumber &=\sum^{N-1}_{n=1} \beta_{t_{n+1}}\tr{\pi(\vec{\Lambda}_{t_{n}})H(\vec{\Lambda}_{t_{n+1}})}-\beta_{t_{n}}\tr{\pi(\vec{\Lambda}_{t_{n}})H(\vec{\lambda}_{t_{n}})}, \\
	\nonumber&=\beta_c\sum^{N-1}_{n=1}\tr{H(\vec{\lambda}_{t_{n+1}})-H(\vec{\lambda}_{t_{n}})\big)\pi(\vec{\Lambda}_{t_{n}})}+(\beta_h-\beta_c)\sum^{N-1}_{n=1}\alpha_{t_{n+1}}\tr{H(\vec{\lambda}_{t_{n+1}})\pi(\vec{\Lambda}_{t_{n}})}-\alpha_{t_{n}}\tr{H(\vec{\lambda}_{t_{n}})\pi(\vec{\Lambda}_{t_{n}})}, \\
	\nonumber&=\beta_c W+(\beta_h-\beta_c)\sum^{N-1}_{n=1}\alpha_{t_{n+1}}\tr{H(\vec{\lambda}_{t_{n+1}})\big(\pi(\vec{\Lambda}_{t_{n}})-\pi(\vec{\Lambda}_{t_{n+1}})\big)}, \\
	&=\beta_c W+(\beta_c-\beta_h)Q_{\text{in}},
\end{align}
where we used $\beta_{t_n}:=\beta_c+(\beta_h-\beta_c)\alpha_{t_n}$ in the third line and the periodic boundary conditions $\vec{\Lambda}_{t_N}=\vec{\Lambda}_{t_1}$ in the fourth line. We now expand the irreversible entropy production (3) using~\eqref{eq:rel_expand}:
\begin{align}
	\nonumber S_{irr}&=\frac{1}{2N}\lim_{N\to\infty}\sum^{N-1}_{n=1} \bigg(\frac{1}{N}\bigg)\int^1_0 ds \ \text{Tr}\bigg( \Delta \Phi_n\pi^s(\vec{\Lambda}_{t_n})\Delta \Phi_n\pi^{1-s}(\vec{\Lambda}_{t_n})\bigg)+\mathcal{O}(1/N^2), \\
	\nonumber&=\frac{1}{2N}\int^1_0 dt \int^1_0 ds  \ \text{Tr}\bigg( \dot{\Phi}(\vec{\Lambda})\pi^s(\vec{\Lambda}_{t})\dot{\Phi}(\vec{\Lambda})\pi^{1-s}(\vec{\Lambda}_{t})\bigg)+\mathcal{O}(1/N^2), \\
	&=\frac{1}{2N}\int^1_0 dt \ \beta^{-1}\int^\beta_0 dx \ \text{Tr}\bigg( \dot{\Phi}(\vec{\Lambda})e^{-x H(\vec{\Lambda}_{t})}\dot{\Phi}(\vec{\Lambda})e^{x H(\vec{\Lambda}_{t})}\pi(\vec{\Lambda}_{t})\bigg)+\mathcal{O}(1/N^2), 
\end{align}
where we used a substitution $x=\beta s$ in the third line. We then use $S_{irr}=\beta_c W+(\beta_c-\beta_h)Q_{\text{in}}$ so that
\begin{align}\label{eq:eta_expand}
	\nonumber\eta&=\eta_C\bigg(1-\frac{S_{irr}}{\beta_c W}\bigg)^{-1}, \\
	&=\eta_C+\frac{\eta_C}{2N\beta_c \mathcal{W}}\int^1_0 dt \ \beta^{-1}\int^\beta_0 dx \ \text{Tr}\bigg( \dot{\Phi}(\vec{\Lambda})e^{-x H(\vec{\Lambda}_{t})}\dot{\Phi}(\vec{\Lambda})e^{x H(\vec{\Lambda}_{t})}\pi(\vec{\Lambda}_{t})\bigg)+\mathcal{O}(1/N^2),
\end{align}
where we used $\mathcal{W}=\lim_{N\to\infty} W$ with
\begin{align}
	\lim_{N\to\infty} W=\lim_{N\to\infty}\sum^{N-1}_{n=1}\bigg(\frac{1}{N}\bigg)\tr{\Delta H_n \pi(\vec{\Lambda}_{t_n})}=\int^1_0 dt \ \tr{\dot{H}(\vec{\lambda}_t)\pi(\vec{\Lambda}_t)}
\end{align}
Finally, we make use of $\dot{\Phi}(\vec{\Lambda})=\dot{\beta}\delta H(\vec{\lambda})+\beta\dot{\lambda}^j \delta X_j$, which upon substitution into~\eqref{eq:eta_expand} yields the metric expression (10).

\section{Example: metric tensor and Pareto fronts for a classical Gaussian system}

\ 

\begin{figure}[t!]
	\includegraphics[width=.5\columnwidth]{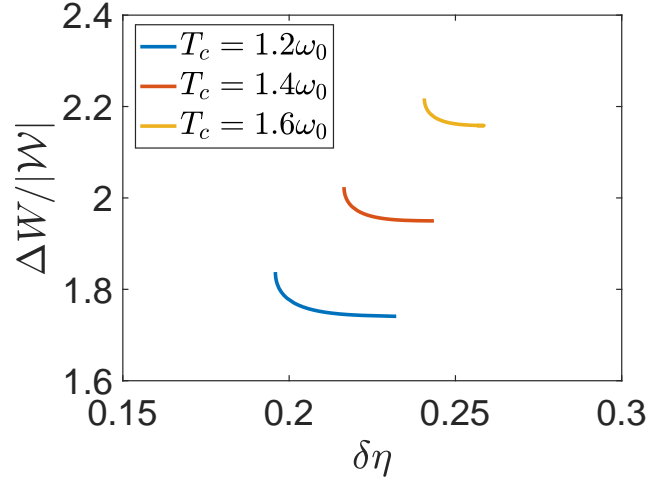}
	\caption{The \textit{relative work-fluctuations vs. efficiency} Pareto fronts for different values of cold temperature $T_c$ for the classical Harmonic oscillator. The parameters are set to  $\omega_0=1$, $T_h = T_c + \Delta T$ with $\Delta T = \omega_0$ and $N=50$.}
	\label{fig:classical_HO}
\end{figure}

We consider here a classical harmonic oscillator system described by a unit mass Hamiltonian $H(z | \omega):=\frac{1}{2}(p^2+\omega^2 x^2)$, where $z=(x,p)$ describes a point in the phase space for the position $x$ and momentum $p$, while $\omega$ is the frequency of the oscillator. Assuming control over the frequency and temperature, we have parameters $\vec{\Lambda}=(T,\omega)$ and a manifold of equilibrium states with probability distribution
\begin{align}
	p(z | \vec{\Lambda}):=\frac{e^{-H(z | \omega)/T}}{Z(\vec{\Lambda})}, \ \ \ \ \ \ Z(\vec{\Lambda})=\int dx \int  dp \ e^{-H(z | \omega)/T},
\end{align}
which may be re-expressed in Gaussian form
\begin{align}
	p(\vec{\Lambda};z)=\frac{e^{-\frac{1}{2}\big(z^{T}\sigma^{-1}(\vec{\Lambda}) \  z\big)}}{2\pi \  \text{det} \ \sigma(\vec{\Lambda})},
\end{align}
with a $2\times2$ positive covariance matrix 
\begin{align}
	\sigma(\vec{\Lambda})=\left[
	\begin{array}{cccc}
		T/\omega^2 & 0  \\
		0 & T \\
	\end{array}
	\right],
\end{align}
For \textit{classical} Gaussian distributions, one may calculate the Fisher-Rao metric tensor~\eqref{eq:fisher} for parameters $\vec{\Lambda}$ as follows:
\begin{align}
	F_{jk}(\vec{\Lambda})&=\frac{1}{2}\text{tr}\bigg(\sigma^{-1}(\vec{\Lambda}) \ \partial_k \sigma(\vec{\Lambda}) \ \sigma^{-1}(\vec{\Lambda}) \  \partial_j \sigma(\vec{\Lambda}) \bigg), 
\end{align}
where $\partial_0=\partial_T$ and $\partial_1=\partial_\omega$. We find
\begin{align}
	&F_{00}(\vec{\Lambda})=\frac{1}{T^2}, \\
	&F_{01}(\vec{\Lambda})=-\frac{1}{\omega  T} \\
	&F_{11}(\vec{\Lambda})=\frac{2}{\omega^2},
\end{align}
We consider a cycle $\gamma: t\mapsto \vec{\Lambda}(t)$ with $t\in[0,1]$ that extracts positive work from the system. The adiabatic work done can be evaluated according to
\begin{align}
	\mathcal{W}=\oint_\gamma \tr{X_j(\vec{\Lambda}) \pi(\vec{\Lambda})}d\lambda^j 
	= \oint_\gamma \omega\tr{x^2 \pi(\vec{\Lambda})}d\omega = \oint_\gamma \omega\frac{T}{\omega^2}d\omega = \oint_\gamma \frac{T}{\omega}d\omega.
\end{align}
Let us consider a cycle of the form
\begin{align}
	&T^{-1}(t)=\beta_c+(\beta_h-\beta_c)\text{sin}^2(\pi t), \\
	&\omega(t)=\omega_0\left(1+\text{sin}^2(\pi t+\frac{\pi}{4})\right).
\end{align}
with the parameters $\omega_0$, $\beta_c$ and $\beta_h$ being fixed during the cycle. As stated in the main text, we can now compute the following tensor:
\begin{align}\label{eq:metric_class}
	M_{jk}^\epsilon(\vec{\Lambda})=\bigg(\epsilon \bigg( \frac{T}{T_c}\bigg)^2 \mu_{jk}+\frac{(1-\epsilon)}{2\beta_c |\mathcal{W}|}\bigg) F_{jk}(\vec{\Lambda}),
\end{align}
with $\mu_{j0}=\mu_{0k}=0 \ \forall j,k$ and $\mu_{jk}=1 \ \forall j,k>0$. To determine the Pareto front for optimal efficiency and fluctuations from this tensor, we follow the steps outlined in the main text and numerically evaluate the speed function $\phi_t^\epsilon$ using Eq. (17). The resulting optimal protocol $\gamma: t\mapsto \vec{\Lambda}'(t)=\vec{\Lambda}(\phi^\epsilon_t)$ provides us with the Pareto front, which we present in Figure~\ref{fig:classical_HO}. These curves represent the boundary of optimal protocols in terms of efficiency and fluctuations for this classical stochastic system, analogous to the quantum mechanical example presented in the main text.

\section{Gaussian formalism for quantum systems}\label{App:Quench}

\

In this section we provide a derivation of the analytic expressions for the metrics  for Gaussian states, namely (19).  Let us start by finding the elements $X_i$ in (2) for the Gaussian scenario. By defining the matrix $\Sigma$ whose elements are the second order symmetric quadratures $\Sigma_{ij} = 1/2\{R_i,R_j\}$ and using the fact that the matrix ${\mathbb G}_{\vec{\lambda}}$ and its derivatives are symmetric, we have
\begin{align}
	X_{j\neq 0} & = \partial_{\vec{\Lambda_{j}}} H({\vec{\lambda}}) =  \frac{1}{2}R^T(\partial_{\vec{\Lambda_{j}}} \mathbb{G}_{\vec{\lambda}})R = \frac{1}{2}\Tr{{ (\partial_{\vec{\Lambda_{j}}} {\mathbb G}_{\vec{\lambda}})^T\Sigma}} \eqqcolon \frac{1}{2}\Tr{{ {\mathbb X}_j \Sigma}},\label{X_j}\\
	\delta X_{j\neq 0} 
	& = \partial_{\vec{\Lambda_{j}}} H({\vec{\lambda}}) - {\rm Tr}\big({\pi(\vec{\Lambda}) \partial_{\vec{\Lambda_{j}}} H({\vec{\lambda}})} \big)
	= \frac{1}{2}\Tr{{ {\mathbb X}_j \Sigma}} - \frac{1}{2}{\rm Tr}\left({\pi(\vec{\Lambda}) \Tr{{ {\mathbb X}_j \Sigma}}} \right) \nonumber\\
	& = \frac{1}{2}\Tr{{ {\mathbb X}_j \Sigma}} - \frac{1}{2}\Tr{ {\mathbb X}_j \sigma(\vec{\Lambda})} = \frac{1}{2}\Tr{{\mathbb X}_j (\Sigma -\sigma(\vec{\Lambda}))}.
\end{align}
Where we defined ${\mathbb{X}_j} \coloneqq \partial_{\vec{\Lambda_{j}}} {\mathbb G}_{\vec{\lambda}}$. Notice that ``$\tr{.}$'' is different from ``$\Tr{}$'', which represents the expectation value.
As for the temperature element $X_0$ we have
\begin{align}
	\delta X_{0} = \beta^{-1} \left(H({\vec{\lambda}}) - {\rm Tr}\left(H({\vec{\lambda}}) \pi({\vec{\Lambda}})\right)\right) 
	& = \frac{\beta^{-1}}{2}\Tr{{\mathbb G}_{\vec{\lambda}} (\Sigma-\sigma(\vec{\Lambda})} \eqqcolon \frac{1}{2}\Tr{{\mathbb X}_{0} (\Sigma-\sigma(\vec{\Lambda}))} 
\end{align}
where we define ${\mathbb X}_{0} \coloneqq \beta^{-1} {\mathbb G}_{\vec{\lambda}}$.

In the Gaussian formalism we know how the displacement vector and the covariance matrix---and hence all higher order moments---evolve under unitary transformations. In particular, under a unitary that is generated by the Hamiltonian $H({\vec{\lambda}}) = 1/2 R^T {\mathbb G}_{\vec{\lambda}} R$ we have
\begin{align}
	\mathscr{U}_{\nu,\vec{\lambda}}\left[\Sigma - \sigma(\vec{\Lambda})\right] &= e^{i \nu H({\vec{\lambda}})}(\Sigma - \sigma(\vec{\Lambda})) e^{-i \nu H({\vec{\lambda}})} = S_{{\mathbb G}_{\vec{\lambda}}}^{\nu}\Sigma \ S_{{\mathbb G}_{\vec{\lambda}}}^{\nu T} - \sigma(\vec{\Lambda}),\\
	\mathscr{U}_{\nu,\vec{\lambda}}\left[R\right] &= e^{i \nu H({\vec{\lambda}})} R \ e^{-i \nu H({\vec{\lambda}})} = S_{{\mathbb G}_{\vec{\lambda}}}^{\nu}R_{\vec{\lambda}}
\end{align}
where we use the BCH lemma, and define $S_{{\mathbb G}_{\vec{\lambda}}}^{\nu}= e^{-i\nu\Omega {\mathbb G}_{\vec{\lambda}}}$. With these at hand we can work on the metric (11)
\begin{align}\label{eq_integrand_xi}
	g_{jk}(\vec{\Lambda}) &= \beta\int^\beta_0 dx \  \tr{\delta X_{k}\mathscr{U}_{ix,\vec{\lambda}}\left[\delta X_{j}\right] \pi(\vec{\Lambda})}, \\ 
	&=\beta\int^\beta_0 dx \  \frac{1}{4}\text{Tr}\Big(\Tr{(\Sigma - \sigma(\vec{\Lambda}) ) {\mathbb X}_k}\nonumber\\
	&\times \Tr {(S_{{\mathbb G}_{\vec{\lambda}}}^{ix}(\Sigma - \sigma(\vec{\Lambda})) ~S_{{\mathbb G}_{\vec{\lambda}}}^{ix T}){\mathbb X}_j}\pi(\vec{\Lambda})\Big)\nonumber\\
	& 
	=\beta\int^\beta_0 dx \ \frac{1}{2}\Tr{(S_{{\mathbb G}_{\vec{\lambda}}}^{ix T} {\mathbb X}_j S_{{\mathbb G}_{\vec{\lambda}}}^{ix} )((\sigma(\vec{\Lambda})-\frac{1}{2}\Omega) {\mathbb X}_k (\sigma(\vec{\Lambda})+\frac{1}{2}\Omega))},
\end{align}
where we used the \textit{Wick's theorem} in order to expand the fourth order correlations in terms of second moments. Recall that we also have the first moments, and hence all odd moments, vanish. In the same manner, we can find the work fluctuations metric: 
\begin{align}
	m_{jk}(\vec{\Lambda}) & = \frac{1}{2}({\tilde m}_{jk}(\vec{\Lambda}) + {\tilde m}_{kj}(\vec{\Lambda})),\nonumber\\
	{\tilde m}_{jk}(\vec{\Lambda}) & = \frac{1}{4}\text{Tr}\big(\Tr{(\Sigma - \sigma(\vec{\Lambda})) {\mathbb X}_k}\Tr{(\Sigma - \sigma(\vec{\Lambda})) {\mathbb X}_j} \pi(\vec{\Lambda})\big), \\
	& = \frac{1}{2}\Tr{ {\mathbb X}_j  \big(\sigma(\vec{\Lambda})-\frac{1}{2}\Omega) {\mathbb X}_k (\sigma(\vec{\Lambda})+\frac{1}{2}\Omega\big)},
\end{align}
which completes the proof of (19).

Finally, if we substitute~\eqref{X_j} in (12) of the main text, we find the adiabatic work in the Gaussian formalism
\begin{align}\label{eq:adaibatic_work_Gaussian}
	\mathcal{W}=\oint_\gamma d\lambda^j  \tr{X_j(\vec{\Lambda}) \pi(\vec{\Lambda})}
	= \frac{1}{2} \oint_\gamma d\lambda^j{\text{Tr}}\left[{\rm tr}\left({\mathbb{X}_j \Sigma}\right) \pi(\vec{\Lambda})\right]
	= \frac{1}{2}\oint_\gamma d\lambda^j{\rm tr}\left({\mathbb{X}_j \sigma(\vec{\Lambda})}\right).
\end{align}
In evaluation of \eqref{eq:adaibatic_work_Gaussian} recall that the temperature element is not included in the integration, i.e., $j\geq 1$.

\section{Thermodynamic geometry for open quantum systems}

\

In this section, we demonstrate how our formalism in the main text can be extended beyond step-equilibration processes to continuous Markovian processes. Rather than modelling the system evolution by a sequence of Hamiltonian quenches followed by relaxation, we instead consider a weakly coupled open system $\rho_t$ whose evolution over time interval $t\in[0,\tau]$ is given by a time-dependent Lindbladian:
\begin{align}
	\dot{\rho}=\mathscr{L}_{\vec{\Lambda}}[\rho],
\end{align}
This Lindbladian depends on both temperature and the mechanical variables of the corresponding Hamiltonian $H(\vec{\lambda})$, collectively labelled by $\vec{\Lambda}=\{\beta,\vec{\lambda}\}$ as before. We assume the evolution obeys quantum detailed balance \cite{Holevo2001} with a unique thermal fixed point for each parameter $\vec{\Lambda}$
\begin{align}
	\mathscr{L}_{\vec{\Lambda}}[\pi(\vec{\Lambda})]=0,
\end{align}
The process is cyclical and described by a closed curve $\gamma:t\mapsto \vec{\Lambda}_t$. In this case, the average work done 
\begin{align}
	W=\int^\tau_0 dt \ \tr{\dot{H}(\vec{\lambda}_t)\rho_t},
\end{align}
and irreversible entropy production
\begin{align}
	S_{irr}:=-\int^\tau_0 dt \ \beta(t) \tr{H(\vec{\lambda}_t)\mathscr{L}_{\vec{\Lambda}_t}[\rho_t]},
\end{align}
As we saw with~\eqref{eq:eta_expand}, the corresponding efficiency of the process can be related to the ratio between entropy production and work \cite{Brandner2018}:
\begin{align}
	\eta=\eta_C\bigg(1-\frac{S_{irr}}{\beta_c W}\bigg)^{-1},
\end{align}
We now restrict to a regime where the system remains close to equilibrium at all times. This occurs whenever the characteristic timescale $t^{eq}$ of the system is always small compared to the total duration, namely $(t^{eq}/\tau)^2\ll 1$ \cite{Cavina2017}. This slow driving regime can be thought of as an analogue of the large step approximation used in the main text. By expanding up to linear order in $t^{eq}/\tau$, it can be shown (see \cite{Brandner2020,Miller}) that the fraction of efficiency below Carnot is
\begin{align}
	\delta \eta:=1-\frac{\eta}{\eta_C}\simeq-\frac{\eta_C}{\tau \beta_c \mathcal{W}}\int_\gamma dt \  \ g_{jk}(\vec{\Lambda}_t)\frac{d\Lambda^j}{dt}\frac{d\Lambda^k}{dt}, 
\end{align}
where $\mathcal{W}$ is the adiabatic work as defined in (12), and we introduce a new metric tensor
\begin{align}\label{eq:metric_g_open_General}
	g_{jk}(\vec{\Lambda}):=\frac{1}{2}\bigg(\tilde{g}_{kj}(\vec{\Lambda})+\tilde{g}_{jk}(\vec{\Lambda})\bigg),
\end{align}
where
\begin{align}
	\tilde{g}_{jk}(\vec{\Lambda}_t):=\beta \int^\beta_0 dx \ \int^\infty_0 d\nu \ \tr{\pi(\vec{\Lambda}) \ e^{\nu\mathscr{L}^\dagger_{\vec{\Lambda}}}\big[\delta X_j(\vec{\lambda})\big] \mathscr{U}_{ix,\vec{\lambda}}\big[\delta X_k(\vec{\Lambda})\big]},
\end{align}
with $\mathscr{L}^\dagger$ the adjoint. 

To quantify the work fluctuations $\text{Var}(W)$, one needs to unravel the system evolution in terms of quantum jump trajectories \cite{Leggio2013a,Horowitz2013b,Manzano2015}. This amounts to monitoring the energy exchanges with the environment, with jump statistics determined by Born's rule. The precise formalism for computing $\text{Var}(W)$ can be found in \cite{Miller}. There it is shown that under slow driving the work fluctuations can be expressed as:
\begin{align}
	\text{Var}(W)=\frac{2}{\tau}\int_\gamma dt \ m_{ij}(\vec{\Lambda}_t)\frac{d\lambda^i}{dt}\frac{d\lambda^j}{dt},
\end{align}
The corresponding metric is given by 
\begin{align}\label{eq:metric_m_open_General}
	m_{jk}(\vec{\Lambda}):=\frac{1}{2}\bigg(\tilde{m}_{jk}(\vec{\Lambda})+\tilde{m}_{kj}(\vec{\Lambda})\bigg)
\end{align}
where
\begin{align}
	\tilde{m}_{jk}(\vec{\Lambda}):=
	\frac{1}{2}\int^\infty_0 d\nu \ \tr{\pi(\vec{\Lambda}) \big\{e^{\nu\mathscr{L}^\dagger_{\vec{\Lambda}}}\big[\delta X_j(\vec{\Lambda})\big], \delta X_k(\vec{\Lambda}) \big\}}, \ \ \ \text{if} \ \ \ \ \ j,k\geq 1
\end{align}
and $\tilde{m}_{j0}=0 \ \ \forall j>0$. With these two metrics, one may minimise the objective function (13) by constructing the analogous metric (15). 

Going further, we now derive a set of closed expressions for evaluation of the thermodynamical quantities in the Gaussian formalism. The main difference with the non-dissipative case---that was presented in the main text and proved in the previous section---is the presence of a Gaussian Lindbladian master equation---which is determined by the term $e^{\nu\mathscr{L}^\dagger_{\vec{\Lambda}}}$. Since the dissipative dynamics is Gaussian too, it can always be characterized efficiently. Generally speaking for an arbitrary observable $O$ we have
\begin{align}
	\mathscr{L}^\dagger_{\vec{\Lambda}} [O] =  i[H_{\vec{\lambda}},O] + \sum_{k=1}^{m}\left(L_{k,{\vec{\Lambda}}}^{\dagger} O L_{k,{\vec{\Lambda}}} - \frac{1}{2}\left\{L_{k,{\vec{\Lambda}}}^{\dagger} L_{k,{\vec{\Lambda}}},O\right\}\right),
\end{align}
with $H_{\vec{\lambda}} = \frac{1}{2} R^T{\mathbb G}_{\vec{\lambda}} R$ being the Hamiltonian,
and the Lindbladian operators $ L_{k,{\vec{\Lambda}}} = c^T_{k,{\vec{\Lambda}}} R$ are linear in quadratures. Here, $c_{k,{\vec{\Lambda}}}$ are $2D$ dimensional complex vectors.

Like any other Gaussian quantuman channel one only needs to identify how they transform the first and second order moments. In our case, we only focus on vanishing first order moments. 
Then the dissipative Gaussian channel can be characterized in the Heisenberg picture as follows
\begin{align}
	e^{\nu\mathscr{L}^\dagger_{\vec{\Lambda}}}\left[\Sigma\right] 
	= F_{\nu,\vec{\Lambda}} \Sigma F_{\nu,\vec{\Lambda}}^T + Y_{\nu,\vec{\Lambda}},
\end{align}
where the matrices $F_{\nu,\vec{\Lambda}}$ and $Y_{\nu,\vec{\Lambda}}$ can be found from the basic elements of the Gaussian Lindbladian master equation i.e, $G_{\vec{\lambda}}$ and $c_{k,\vec{\Lambda}}$. Specifically, we have 
$F_{\nu,\vec{\Lambda}} = e^{\nu A_{\vec{\Lambda}}}$ and $Y_{\nu,\vec{\Lambda}} = \int_0^{\nu} d\nu^{\prime} e^{\nu^{\prime} A_{\vec{\Lambda}}} D_{\vec{\Lambda}} e^{\nu^{\prime} A_{\vec{\Lambda}}^{T}}$
with $A_{\vec{\Lambda}} = -i\Omega({\mathbb G}_{\vec{\lambda}} - {\rm Im}(C_{\vec{\Lambda}} C_{\vec{\Lambda}}^{\dagger}))$, and $D_{\vec{\Lambda}} = \Omega~{\rm Re}(C_{\vec{\Lambda}} C_{\vec{\Lambda}}^{\dagger}) \Omega$, where we define $C \coloneqq (c_1^T; c_2^T; \dots c_m^T)^T $---see e.g., Section 5 and Appendix C of \cite{Mehboudi} for derivation.
The application of the channel to $\sigma(\vec{\Lambda})$ should be understood through its application on the identity operator, because in fact by $[\sigma(\vec{\Lambda})]_{jk}$ we mean $[\sigma(\vec{\Lambda})]_{jk} {\mathbb I}$. Since, the map is unital, it leaves $\sigma(\vec{\Lambda})$ unchanged.  
Thus, when applied to $\Sigma-\sigma(\vec{\Lambda})$ we have
\begin{align}
	e^{\nu\mathscr{L}^\dagger_{\vec{\Lambda}}}\left[\Sigma - \sigma(\vec{\Lambda})\right] = F_{\nu,\vec{\Lambda}} \Sigma F_{\nu,\vec{\Lambda}}^T  + Y_{\nu,\vec{\Lambda}} - \sigma(\vec{\Lambda}) = F_{\nu,\vec{\Lambda}} (\Sigma - \sigma(\vec{\Lambda})) ~ F_{\nu,\vec{\Lambda}}^T,
\end{align}
where we use the fact that $\sigma(\vec{\Lambda})$ is the fixed point of the dissipative dynamics i.e., $\sigma(\vec{\Lambda}) =  F_{\nu,\vec{\Lambda}} \sigma(\vec{\Lambda}) ~ F_{\nu,\vec{\Lambda}}^T + Y_{\nu,\vec{\Lambda}}$ for $\forall \nu$. 

Putting everything together we can find the metrics.~\eqref{eq:metric_m_open_General} and \eqref{eq:metric_g_open_General}. Firstly, we have
\begin{align}
	\tilde{m}_{jk}(\vec{\Lambda})& = 
	\frac{1}{2}\int^\infty_0 d\nu \ \tr{\pi(\vec{\Lambda}) \big\{e^{\nu\mathscr{L}^\dagger_{\vec{\Lambda}}}\big[\delta X_j(\vec{\Lambda})\big], \delta X_k(\vec{\Lambda}) \big\}}\nonumber\\
	& = \frac{1}{4}{\rm Re}\int^\infty_0 d\nu \ {\text{Tr}}\bigg(\Tr{(F_{\nu,\vec{\Lambda}}(\Sigma - \sigma(\vec{\Lambda})) F_{\nu,\vec{\Lambda}}^T) {\mathbb X}_k} 
	{\rm tr}\big( (\Sigma - \sigma(\vec{\Lambda})) {\mathbb X}_j\big) \pi(\vec{\Lambda})\bigg), \\
	& = \frac{1}{2}{\rm Re}\int^\infty_0 d\nu \ \Tr{ {\mathbb X}_j  \big((\sigma(\vec{\Lambda})-\frac{1}{2}\Omega) F_{\nu,\vec{\Lambda}}^T {\mathbb X}_k F_{\nu,\vec{\Lambda}}(\sigma(\vec{\Lambda})+\frac{1}{2}\Omega)\big)}\nonumber\\
	&
	= \frac{1}{2} {\rm Re}~{\rm tr}({\mathbb X}_j {\underline {\mathbb X}_k})
	,
\end{align}
where we used the \textit{Wick's theorem} in order to expand the fourth order correlations in terms of second moments. We also extend the definition of ${\underline {\mathbb X}_j}$---from (21)---to the open dynamic scenario
\begin{align}\label{eq:underline_open}
	{\underline {\mathbb X}_j} & = (\sigma(\vec{\Lambda})-\frac{1}{2}\Omega) \left[\int^\infty_0 d\nu \ F_{\nu,\vec{\Lambda}}^T {\mathbb X}_j F_{\nu,\vec{\Lambda}}\right] (\sigma(\vec{\Lambda})+\frac{1}{2}\Omega).
\end{align}
Moreover, the matrix $\tilde{g}_{jk}(\vec{\Lambda})$ can be found in a similar manner:
\begin{align}\label{eq_integrand_xi}
	\tilde{g}_{jk}(\vec{\Lambda}_t)&=\beta \int^\beta_0 dx \ \tr{\pi(\vec{\Lambda}) \ e^{\nu\mathscr{L}^\dagger_{\vec{\Lambda}}}\big[\delta X_j(\vec{\lambda})\big] \mathscr{U}_{ix,\vec{\lambda}}\big[\delta X_k(\vec{\Lambda})\big]},\nonumber\\
	&=\frac{\beta}{4}\int^\infty_0 d\nu \int^\beta_0 dx \  {\text{Tr}}\Big({\rm tr}\big(F_{\nu,\vec{\Lambda}}(\Sigma - \sigma(\vec{\Lambda})) F_{\nu,\vec{\Lambda}}^T {\mathbb X}_k\big){\rm tr} \big(S_{{\mathbb G}_{\vec{\lambda}}}^{ix}(\Sigma - \sigma(\vec{\Lambda})) ~S_{{\mathbb G}_{\vec{\lambda}}}^{ix T}{\mathbb X}_j \big)\pi(\vec{\Lambda})\Big)\nonumber\\
	& 
	=\frac{\beta}{2}\int^\infty_0 d\nu \int^\beta_0 dx \  {\rm tr}\big((S_{{\mathbb G}_{\vec{\lambda}}}^{ix T} {\mathbb X}_j S_{{\mathbb G}_{\vec{\lambda}}}^{ix} )((\sigma(\vec{\Lambda})-\frac{1}{2}\Omega) F_{\nu,\vec{\Lambda}}^T {\mathbb X}_k F_{\nu,\vec{\Lambda}}(\sigma(\vec{\Lambda})+\frac{1}{2}\Omega))\big)\nonumber\\
	& = \frac{\beta}{2} {\rm tr} ({\bar {\mathbb X}_j} {\underline {\mathbb X}_k}),
\end{align}
where ${\bar {\mathbb X}_j}$ is still given by (20) whereas $\underline {\mathbb X}_k$ is given by \eqref{eq:underline_open} above.

\end{document}